\documentclass[
reprint,
superscriptaddress,
amsmath,amssymb,
aps,
]{revtex4-2}
\usepackage{graphicx}
\usepackage{dcolumn}
\usepackage{bm}
\usepackage{float}
\usepackage{multirow}
\usepackage{xcolor}
\usepackage{hyperref}
\hypersetup{
colorlinks = true,	
urlcolor = blue,	
linkcolor= blue,	
citecolor = blue 
}

\begin{document}


\title{Emergence of order from chaos through a continuous phase transition in a turbulent reactive flow system}

\author{Sivakumar Sudarsanan}
\affiliation{%
 Department of Aerospace, Indian Institute of Technology Madras, Chennai  600 036, India
}%
\affiliation{Centre of Excellence for studying Critical Transition in Complex Systems, Indian Institute of Technology Madras, Chennai  600 036, India}

\author{Amitesh Roy}%
\affiliation{Institute for Aerospace Studies, University of Toronto, Ontario, Canada M3H 5T6}

\author{Induja Pavithran}%
\author{Shruti Tandon}%
\author{R. I. Sujith}%
\email{sujith@iitm.ac.in}
\affiliation{%
 Department of Aerospace, Indian Institute of Technology Madras, Chennai  600 036, India
}%
\affiliation{Centre of Excellence for studying Critical Transition in Complex Systems, Indian Institute of Technology Madras, Chennai  600 036, India}  

\date{\today}

\begin{abstract}
As the Reynolds number is increased, a laminar fluid flow becomes turbulent, and the range of time and length scales associated with the flow increases. Yet, in a turbulent reactive flow system, as we increase the Reynolds number, we observe the emergence of a single dominant time scale in the acoustic pressure fluctuations, as indicated by its loss of multifractality. Such emergence of order from chaos is intriguing. We perform experiments in a turbulent reactive flow system consisting of flame, acoustic, and hydrodynamic subsystems interacting nonlinearly. We study the evolution of short-time correlated dynamics between the acoustic field and the flame in the spatiotemporal domain of the system. The order parameter, defined as the fraction of the correlated dynamics, increases gradually from zero to one. We find that the susceptibility of the order parameter, correlation length, and correlation time diverge at a critical point between chaos and order. Our results show that the observed emergence of order from chaos  is a continuous phase transition. Moreover, we provide experimental evidence that the critical exponents characterizing this transition fall in the universality class of directed percolation. Our study demonstrates how a real-world complex, nonequilibrium turbulent reactive flow system exhibits universal behavior near a critical point.
\end{abstract}

\maketitle

\section{\label{sec:level1}INTRODUCTION\protect}
The spontaneous emergence of order from a chaotic turbulent state in the form of self-sustained periodic oscillations is encountered frequently in nonequilibrium systems. Examples of such spontaneous oscillations include birdsong, whistling, and the sounds of woodwinds, which are very pleasant \cite{fee1998role,anderson1952dependence,hirschberg1989whistler,howe1975contributions}. However, 
within engineering systems, uncontrolled oscillations are highly undesirable as they can frequently result in catastrophic failures of gas transport pipelines \cite{kriesels1995high}, rockets and gas turbine engines \cite{fisher2009nasa,lieuwen2005combustion}, and bridges, for example, the devastating collapse of the Tacoma bridge \cite{green2006failure}. 

As the Reynolds number (which is the ratio of inertial to viscous force) is increased, the range of time and length scales associated with a fluid flow increases \cite{pope2001turbulent, davidson2015turbulence}. Yet, in a turbulent reactive flow system, as we increase the Reynolds number, we observe the emergence of a single dominant time scale in the acoustic pressure fluctuations, as indicated by its loss of multifractality \cite{nair2014multifractality}. In this work, we study the emergence of order from chaos in a turbulent thermo-fluid system. 

Recent studies in disparate turbulent flow systems reported universal scaling relations during the transition from chaos to order \cite{pavithran2020universality_oscillatory,pavithran2020universality_spectral}. Such universal scaling laws observed during a transition are often associated with a second-order phase transition and characterized by critical exponents \cite{Hinrichsen2000,tauber2017phase}. Systems that exhibit the same critical exponents are identified with a universality class. The transition in fluid flows from a laminar to a turbulent state follows the universality class of directed percolation (DP) \cite{sipos2011directed,hof2023directed,lemoult2016directed,moxey2010distinct}. Moreover, the synchronization transition in coupled map lattices \cite{ginelli2003relationship,baroni2001transition} and cellular automata \cite{grassberger1999synchronization}, transition in turbulent liquid crystals \cite{takeuchi2007directed}, and numerous other phenomena \cite{chantelot2021leidenfrost,Hinrichsen2000,shrivastav2016yielding,rupp2003critical} fall under the universality class of DP.  

In the present work, we perform experiments in a confined turbulent reactive flow system, which consists of flame, hydrodynamic, and acoustic subsystems interacting nonlinearly. As the Reynolds number $\mathrm{Re}$ is increased, self-sustained periodic oscillations of large magnitude emerge in the acoustic pressure as a result of a positive feedback between the different subsystems \cite{sujith2020complex, angeli2004detection}. This phenomenon, referred to as thermoacoustic instability or combustion instability is a significant concern for gas turbine power plants, aircraft and rocket engines due to the catastrophic consequences associated with high amplitude acoustic pressure oscillations \cite{fisher2009nasa}. 

We study the evolution of short-time correlated dynamics between the acoustic field and the flame in the spatiotemporal domain of a turbulent reactive flow system. We show that during the emergence of order from a chaotic state, the order parameter, defined as the fraction of correlated dynamics, increases gradually from zero to one. Close to the onset of the transition (critical point), the fluctuations in the order parameter become significant and the variance of the order parameter fluctuations, referred to as susceptibility, exhibits a diverging behavior. We find that the correlations in fluctuations persist longer in the vicinity of the critical point. {In particular, the correlation length and correlation time diverge according to a power law at the critical point. These measures clearly imply the occurrence of a continuous phase transition from a chaotic to an ordered state even as the Reynolds number is increased.} Further, we find that three critical exponents of the transition corresponding to the order parameter and the distribution of intervals between the occurrence of correlated dynamics along space and time fall into the universality class of 2+1 DP. \textcolor{black}{In summary, we show that a highly turbulent nonlinear system follows universal scaling laws close to the critical point of a continuous phase transition.} 

\section{\label{sec:level2}PHASE TRANSITION IN NONEQUILIBRIUM SYSTEMS\protect}

In complex systems, a phase represents the collective state of multiple interacting subsystems. When a suitable control parameter is varied, complex systems can exhibit a qualitative change in the collective state of the system. In nonequilibrium systems, these collective, self-organized states emerge and are maintained due to the constant flux of energy \cite{prigogine1978time}. Examples of phase transition in nonequilibrium systems include sleep-wake transition \cite{henn1984sleep}, the emergence of coherent light emission in lasers \cite{haken1977synergetics}, \textcolor{black}{the transition from order to chaos occurring in transitional turbulent flows such as Rayleigh-B\'{e}nard convection \cite{eckhardt2018transition, ecke1984critical}, Couette flow \cite{eckhardt2018transition, kreilos2012periodic}, and turbulent swirling flow with a pair of counter-rotating impellers \cite{faranda2017stochastic}.}

DP model typifies phase transition in diverse nonequilibrium systems \cite{Hinrichsen2000}. This model describes the spread of activity through contact processes, such as the spreading
of an epidemic in a community or the spread of forest fires \cite{Hinrichsen2000}. If the spreading probability $p$ is less than a critical value ($p_c$), the system evolves to a state with no activity in the system dynamics. As the spreading probability increases, for $p>p_c$, the activity exhibits a percolation phase transition \cite{Hinrichsen2000}.

Order parameter ($\rho$) quantifies the density of active sites in the system during the stationary state. During the DP phase transition, the system exhibits critical scaling behavior near the critical point, $p_c$ \cite{Hinrichsen2000,henkel2008non}. The order parameter exhibits a power law $\rho \sim \epsilon^\beta$ with an exponent $\beta$, where $\epsilon=p-p_c$ is the distance from the critical point ($p_c$).

Let the interval between the consecutive active sites in space be $\ell$, at a time instant, and the duration between the consecutive active instances be $\tau$ at a spatial location. These intervals between the occurrences of activity are defined as inactive intervals. The probability distributions ($\mathcal{N}$) of inactive time intervals ($\tau$) and length intervals ($\ell$) exhibit power law with $\mathcal{N}(\ell) \sim \ell^{-\mu_{\parallel}}$ and  $\mathcal{N}(\tau) \sim \tau^{-\mu_{\perp}}$ at the critical point of the DP phase transition \cite{Hinrichsen2000}. The critical behavior associated with the universality class of DP is characterized by power law exponents $\beta$, $\mu_\parallel$, and $\mu_\perp$ \cite{Hinrichsen2000, hinrichsen2000possible}. Diverse systems including fluid \cite{sano2016universal, lemoult2016directed, chantelot2021leidenfrost}, material \cite{shrivastav2016yielding, buldyrev1992anomalous}  and biological systems \cite{carvalho2021subsampled} exhibit the same critical behavior as DP. 

\section{\label{sec:level3}Experiments in a Turbulent reactive flow system}
\begin{figure}[ht]
    \centering
    \includegraphics[width=\linewidth]{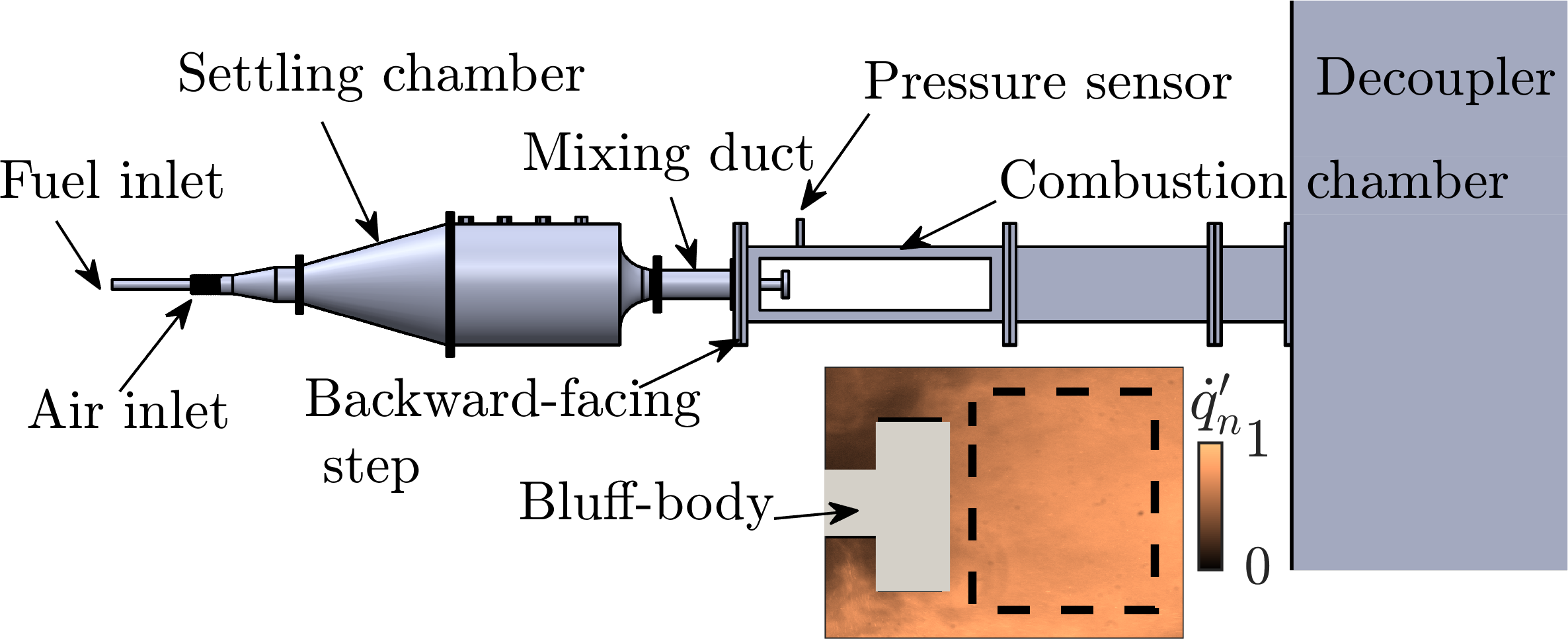}  
    \caption{The schematic of a bluff body stabilized turbulent reactive flow system comprising fuel and air inlets,  a settling chamber, and a combustion chamber. The acoustic pressure is measured using a piezoelectric transducer mounted on the combustion chamber while the heat release rate is measured using CH$^*$ chemiluminescence imaging using a high-speed camera, whose field of view is shown in the inset. The spatial region indicated by the dashed line is used for further analysis.}
    \label{fig: experimental set up}
\end{figure}

Our expt. setup, the turbulent reactive flow system, consists of a mixing duct, bluff body, combustion chamber, and a settling chamber (Fig.~\ref{fig: experimental set up}). Air is first passed through a settling chamber to reduce fluctuations from the air supply line.  The fuel (liquified petroleum gas with a composition of $60\%$ butane and $40\%$ propane) is supplied into the mixing tube through radial injection holes of the central shaft, then mixes with the air from the settling chamber and flows into the combustion chamber. The combustion chamber is a duct of length of 1100 mm and with a square cross-section of 90 mm $\times$ 90 mm. One side of the combustion chamber is a backward-facing step through which the reactant mixture enters the combustion chamber. The reactant mixture is ignited using a spark plug attached to the backward-facing step of the combustion chamber.  The other end of the combustion chamber is connected to a rectangular chamber (or decoupler) of size 1000 mm $\times$ 500 mm $\times$ 500 mm, much larger than its cross-section. This rectangular chamber is to isolate the combustion chamber from external ambient fluctuations. For optical access to the combustion chamber, two quartz glass windows of 90 mm $\times$ 360 mm are provided on both side walls. 

The fundamental mode of the combustion chamber is excited during the emergence of periodic oscillations with a frequency of 160 Hz ($f=c/4L$, $c$ is the speed of sound, and $L$ is the length of the combustor). A circular bluff body of diameter 47 mm and thickness 10 mm fixed at a location 32 mm from the backward-facing step creates a wake flow where the flame is stabilized. The flow rates of air and fuel are separately controlled by mass flow controllers (Alicat scientific MCR series) with an uncertainty of $\pm 0.8\%$  percent of measured reading + $\pm 0.2 \%$ of full-scale reading.  

\begin{figure*}[ht]
    \centering
    \includegraphics[width=0.9\linewidth]{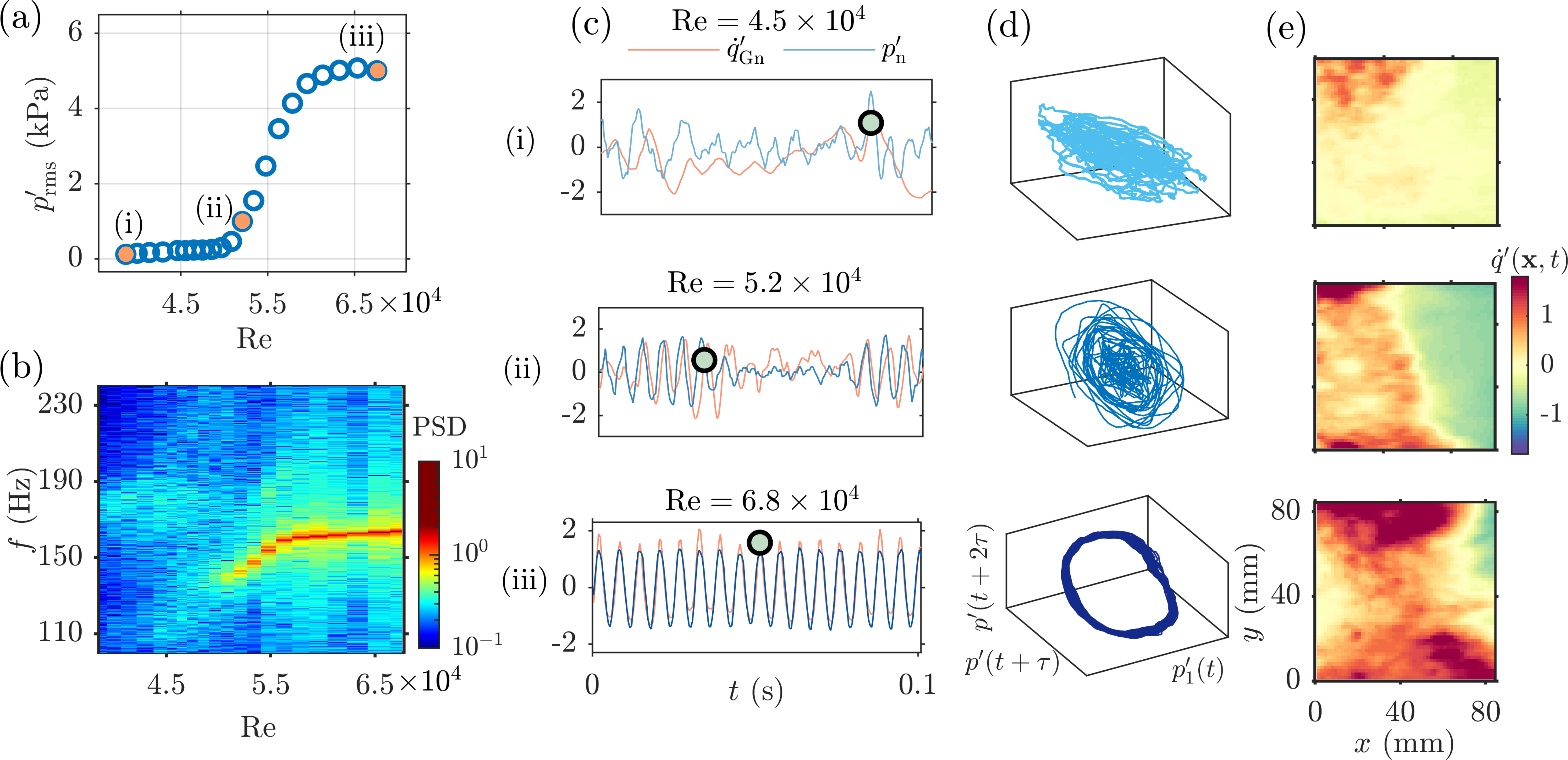}  
    \caption{Transition from chaos to order. (a) The amplitude ($p^\prime_{\text{rms}}$) and (b) the power spectral density (PSD) of $p^\prime$ as a function of $\mathrm{Re}$. With an increase in $\mathrm{Re}$, there is a gradual increase in the amplitude of oscillations accompanied by the appearance of a dominant frequency in the power spectral density of acoustic pressure oscillations. Representative (c) time series (normalized $p^\prime$ and $\dot{q}^\prime_{\mathrm{G}}$) and (d) phase space trajectory for $p^\prime$ corresponding to the dynamical states of chaos, intermittency and limit cycle oscillations, indicated in (a). The phase space is reconstructed using time delay ($\tau$) embedding. (e) The distribution of heat heat release rate fluctuation $\dot{q}^\prime(\textbf{x},t)$ at the indicated instances in (c).}
    \label{fig: transition}
\end{figure*}

We fix the mass flow rate of fuel at 1.75 g/s and vary the mass flow rate of air from 15.3 g/s to 29.5 g/s. This results in the variation in Reynolds number from $\mathrm{Re}=3.9 - 6.8 \times 10^4$.
Here, the Reynolds number is defined as $\mathrm{Re}=\rho \bar{v} D/ \mu$, where $\bar{v}$ is the average velocity of the fuel-air mixture entering the combustion chamber, $D$ is the diameter of the bluff body, $\rho$  and $\mu$  are the density and dynamic viscosity of the mixture calculated by considering the variation in the mixture composition as the control parameter is varied \cite{bird1961transport}. The maximum uncertainty in $\mathrm{Re}$ (calculated based on the uncertainty of the mass flow controllers) is $\pm 308$. {The experiments at each of the specified values of $\mathrm{Re}$ are repeated ten times.}

In order to study the emergence of order from chaos, we measure the acoustic pressure fluctuations ($p^\prime (t)$) and the spatial distribution of heat release rate fluctuations ($\dot{q}^{\prime}(\mathbf{x},t)$) inside a rectangular region (Fig.~\ref{fig: experimental set up}, \ref{fig: transition}(e)) of the combustion chamber. The acoustic pressure fluctuations inside the combustion chamber are measured using a piezoelectric pressure transducer (PCB 103B02) mounted 120 mm from the backward-facing step of the combustion chamber. The pressure transducer is mounted to the wall of the combustion chamber using a T-joint mount. A semi-infinite waveguide of length 10 m and inner diameter 4 mm is connected to the transducer mount to minimize the frequency response of the probe. The pressure transducer has a sensitivity of 223.4 mV/kPa and an uncertainty of $\pm 0.15$ Pa. The acoustic pressure is measured with a sampling frequency of 10 kHz for a duration of 3 s. The signals from the piezoelectric pressure transducer were recorded using a data acquisition system (NI DAQ-6346).

The chemiluminescence intensity represents the line-of-sight integrated heat release rate distribution \cite{hardalupas2004local}. The heat release rate fluctuations are determined from the chemiluminescence images acquired using the high-speed Phantom V12.1 camera outfitted with a CH* filter (a narrow band filter of peak at 435 nm with 10 nm FWHM)  along with a 100 mm Carl-Zeiss lens. A region spanning 80 $\times$ 80 $\mathrm{mm}^2$ is imaged at a resolution of 520 $\times$ 520 pixels at a sampling rate of 2000 Hz simultaneously with the acoustic pressure measurements. We perform a coarse-graining operation by combining 10 $\times$ 10 pixels of the chemiluminescence images to decrease noise effects.

\section{\label{sec:level4}Emergence of order from chaos\protect}
Figure \ref{fig: transition} shows the transition in the dynamics of the turbulent reactive flow system when the Reynolds number is increased from $\mathrm{Re}=3.9\times 10^4$ to $\mathrm{Re}=6.8\times 10^4$. Figures \ref{fig: transition}(a) and \ref{fig: transition}(b) depict the change in the root-mean-square (r.m.s.) of the acoustic pressure fluctuations and the corresponding power spectral density, respectively. The low-amplitude, aperiodic oscillations  (Fig.\ref{fig: transition}(c)-i) are identified as high-dimensional chaos \cite{tony2015detecting}. As $\mathrm{Re}$ is increased, we notice that the turbulent reactive flow system undergoes a transition from a state of low-amplitude, high-dimensional chaos to a state characterized by high-amplitude periodic acoustic pressure fluctuations \textcolor{black}{through a state of intermittency where bursts of high-amplitude periodic acoustic pressure fluctuations appear amidst epochs of low-amplitude aperiodic acoustic pressure fluctuations} (Fig.~\ref{fig: transition}(a),  Fig.~\ref{fig: transition}(c)-i, ii, iii).  {The power spectrum, which is broad-band, becomes progressively narrower as $\mathrm{Re}$ is increased.} At the onset of sustained periodic oscillations, the frequency of the dominant mode of acoustic pressure oscillations is 160 Hz, evident in the power spectrum shown in Fig.~\ref{fig: transition}(b). 

{The phase space trajectories associated with the acoustic pressure fluctuations are reconstructed using Takens' embedding theorem \cite{takens1981lecture} and are shown in figure \ref{fig: transition}(d). \textcolor{black}{The optimum time delay is selected as the first local minima of the average mutual information \cite{fraser1986independent} and the suitable embedding dimension is calculated using the false nearest neighbor method \cite{cao1997practical}. The optimum time delay obtained for the states of chaos, intermittency, and periodic oscillations are 1.6, 1.5, and 1.6 ms respectively, with corresponding suitable embedding dimensions of 9, 6, and 5 (refer to Appendix \ref{Appendix: Phase space reconstruction} for more details).} Corresponding to the state of chaotic fluctuations, the trajectory (Fig.~\ref{fig: transition}(d)-i) appears to be cluttered with no clearly defined attractor. During the chaotic state, the trajectory of the system often switches between multiple unstable periodic orbits (UPOs), as the trajectory is ejected along the unstable manifold from one UPO and attracted towards the stable manifold of another UPO \cite{auerbach1987exploring,tandon2021condensation}. The phase space of the chaotic state thus appears haphazard due to the switching between these UPOs (Fig.~\ref{fig: transition}(d)-i).}

Upon further increase in $\mathrm{Re}$ beyond $5.1 \times 10^4$, we observe the state of intermittency (Fig.~\ref{fig: transition}(c)-ii) \cite{nair2014intermittency}. During the state of intermittency, the trajectory of the system transits between an inner chaotic region and outer periodic orbits in the phase space (Fig.~\ref{fig: transition}(d)-ii) \cite{tandon2021condensation}. We observe sustained periodic acoustic pressure oscillations (Fig.~\ref{fig: transition}(c)-iii) as we further increase $\mathrm{Re}$ to $6.8 \times 10^4$. During the emergence of periodic oscillations, the number of unstable periodic orbits decreases and their stability increases. Eventually, a single stable periodic orbit emerges \cite{tandon2021condensation}, and the system exhibits limit cycle oscillations (Fig.~\ref{fig: transition}(d)-iii). 

To investigate the coupled dynamics of the flame and the acoustic pressure, the time series of heat release rate fluctuation is obtained from the chemiluminescence imaging such that $\dot{q}^\prime(\mathbf{x},t)= \dot{q}(\mathbf{x},t)- \overline{\dot{q}}(\mathbf{x})$ where $\overline{\dot{q}}(\mathbf{x})$ is the temporal average, calculated as $1/N \sum_t \dot{q}(\mathbf{x},t)$ with $N$ denoting the number of data points in the time series. The aggregate heat release rate fluctuation is then obtained as $\dot{q}_{\mathrm{G}}^\prime(t)= \sum_{\mathbf{x}} \dot{q}^\prime(\mathbf{x},t)$. The normalized aggregate heat release rate fluctuation ($\dot{q}^\prime_{\mathrm{Gn}}$) is overlaid on normalized $p^\prime$ to observe their relative time evolution as shown in Fig.~\ref{fig: transition}(c). We normalize $\dot{q}^\prime_{\mathrm{G}}$ and $p^\prime$ using their standard deviation values. {Further, in figures \ref{fig: transition}(e)-i, \ref{fig: transition}(e)-ii, and \ref{fig: transition}(e)-iii, the heat release rate fluctuations across the spatial field $\dot{q}^\prime(\textbf{x},t)$ are shown for the instances marked in Fig.~\ref{fig: transition}(c)-i, \ref{fig: transition}(c)-ii, and \ref{fig: transition}(c)-iii respectively}. The flame and acoustic subsystems exhibit desynchronized dynamics (Fig.~\ref{fig: transition}(c)-i) and the values of heat release rate fluctuations remain very small (Fig.~\ref{fig: transition}(e)-i) during the state of chaos. The flame and the acoustic field exhibit synchronized dynamics when the turbulent reactive flow system exhibits sustained periodic limit cycle oscillations (Fig.~\ref{fig: transition}(c)-iii) \cite{pawar2017thermoacoustic, pawar2019temporal, mondal2017onset}. During this state, the heat release rate fluctuations have large values distributed throughout the reaction field (Fig.~\ref{fig: transition}(e)-iii).  The heat release rate distribution for the state of intermittency is shown in figure \ref{fig: transition}(e)-ii. In a previous study, \citet{mondal2017onset} showed that synchronous and asynchronous dynamics between $p^\prime$ and $\dot{q}^\prime(\mathbf{x},t)$ co-exist during the state of intermittency. {Thus, the turbulent reactive flow system clearly shows the emergence of periodic oscillations from a chaotic state through intermittency while we increase the Reynolds number.}  

 \begin{figure*}[ht]
    \centering
    \includegraphics[width=0.85\linewidth]{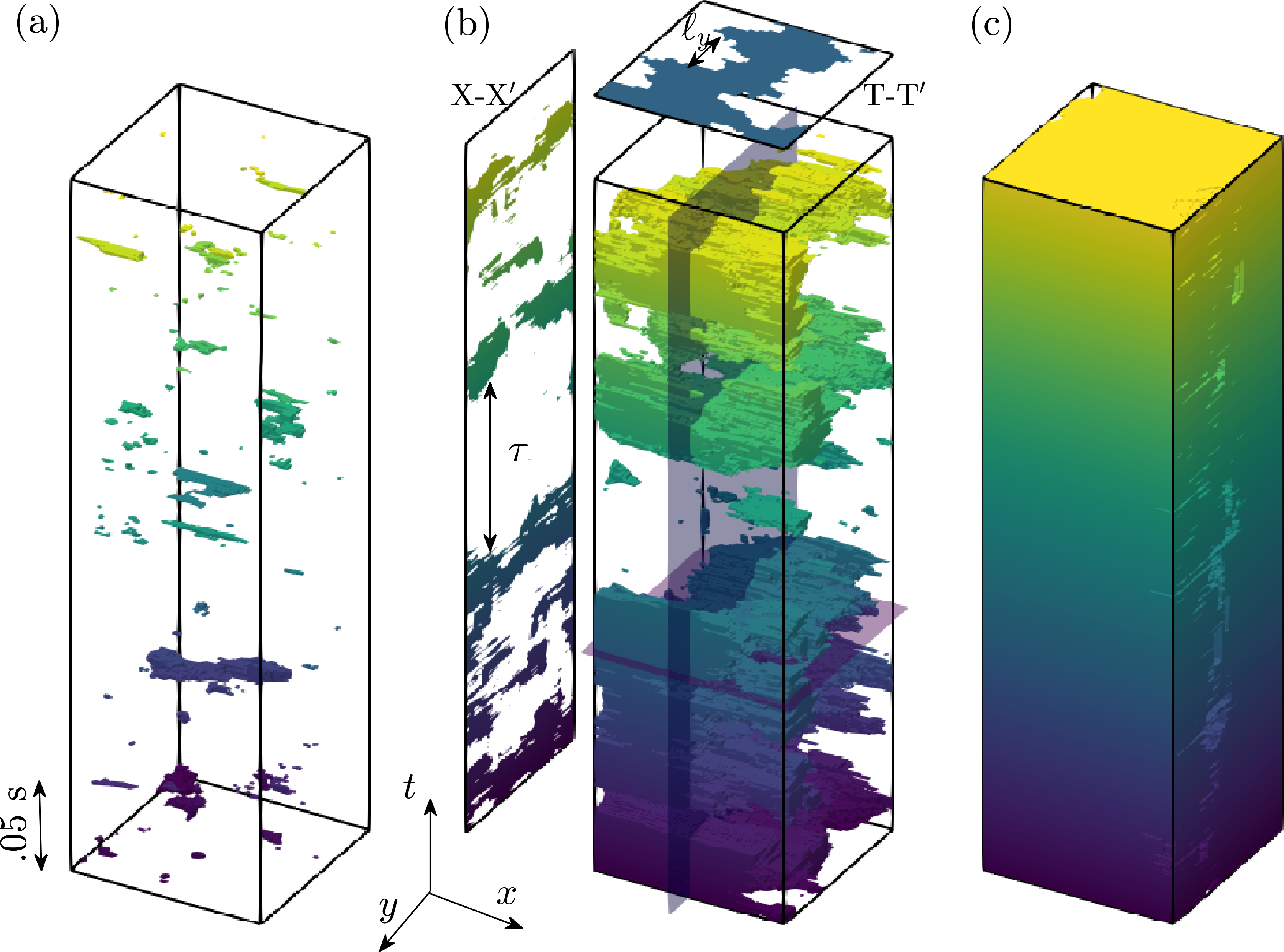}
    \caption{{Space-time ($\textbf{x}-t$) diagram illustrating the phase transition from chaos to order. Panels (a-c) correspond to the state of chaos, intermittency and limit cycle oscillations respectively, as indicated in Fig.~\ref{fig: transition}a}. Regions of ordered activity, identifying the existence of high correlation between the acoustic pressure and the heat release rate fluctuations, is visualized with the color scheme and is delineated from regions of disordered activity using equation \eqref{Eq-StateVariable}. For the chaotic state (a), {the regions of ordered activity are very small and disconnected}. At the critical point associated with the state of intermittency (b), small ordered regions grow and form giant ordered regions. The slice in $X-X^\prime $ and $T-T^\prime$ shows the appearance and disappearance of giant clusters in space and time. {Inactive intervals are labeled as $\ell_y$ in space and $\tau$ in time.} During the state of limit cycle oscillations (c), order percolates throughout the spatiotemporal domain.}
    \label{Fig: percolation}
\end{figure*}

\section{\label{sec:level5}Emergence of order through a continuous phase transition}
We quantify the transition by determining the short window time-delayed cross-correlation between $p^\prime(t)$ and $\dot{q}^\prime(\mathbf{x},t)$. The acoustic pressure fluctuations remain spatially uniform within the region of interest (Fig.~\ref{fig: experimental set up}) since the reaction zone is small in comparison to the acoustic length scales which are of the same order as the length of the combustion chamber (refer to Appendix \ref{Appendix: pressure variation}). Further, since the transition occurs through the state of intermittency which features localized bursts of periodic fluctuations, a short time window is used to capture the intermittent features. The cross-correlation is defined as
\begin{eqnarray}
\mathcal{R}(\mathbf{x},t,\tau_i)={\int_{t}^{t+W} p_{\mathrm{n}}^\prime(t_1)\dot{q}_{\mathrm{n}}^\prime(\mathbf{x},t_1+\tau_i)dt_1}.
\label{eq: corr}
\end{eqnarray}
Here, the correlation is calculated over a short time window of time $W,$ for different time shift values ranging from $\tau_i=-T/2$ to $+T/2$, where $T$ is the time period of the dominant mode of the acoustic pressure oscillations and the subscript `n' signifies that the variables are normalized with their respective standard deviation values. If $W$ is too short, the correlation becomes meaningless and depicts a significant level of fluctuations, whereas if $W$ is too large, the correlation fails to distinguish short bursts of periodic oscillations. Here, we use $W=4T$, where $T=6.25 \times 10^{-3}$ s.

Finally, the short window time-delayed correlation is obtained as
\begin{eqnarray}
R(\textbf{x},t)=\mathrm{max}_{\tau_i}\{\mathcal{R}(\textbf{x},t,\tau_i)\}.
\label{eq: max corr}
\end{eqnarray}
For aperiodic, uncorrelated fluctuations, the cross-correlation values ($\mathcal{R}$) are very small irrespective of the time shift (Fig.~\ref{Fig: Permutation}a). On the other hand, if both the time series are periodic or correlated, the cross-correlation value ($\mathcal{R}$) attains a high value for a time shift value corresponding to the phase difference between the two time series (Fig.~\ref{Fig: Permutation}c).

We introduce the state variable ($S$), defined as
\begin{equation}
  S({\textbf{x},t}) =
    \begin{cases}
      1 & \text{if $R\geq R_\text{Th}$,}\\
      0 & \text{otherwise,} 
    \end{cases}       
    \label{Eq-StateVariable}
\end{equation}
to distinguish between correlated and uncorrelated dynamics. \textcolor{black}{The threshold value, $R_{\mathrm{Th}}=0.6$ is obtained from a surrogate test (Refer to Appendix \ref{Appendix: effect of threshold} for more details). The same value of the threshold, $R_{\mathrm{Th}}=0.6$ is set for all the dynamical states.} 


The scalar variable $S$ classifies the existence of ordered or disordered activity at a given space-time location ($\textbf{x}-t$). 
\textcolor{black}{Thus, $S=0$ represents a disordered activity indicating the presence of uncorrelated dynamics due to turbulence, whereas $S=1$ represents an ordered activity in the form of highly correlated periodic fluctuations.}

\textcolor{black}{The choice of a measure based on linear correlation in our analysis was inspired from the extended Rayleigh criteria for the emergence of self-sustained acoustic pressure oscillations in reactive flow systems \cite{poinsot2005theoretical}. The thermo-acoustic driving is proportional to the linear correlation between the acoustic pressure and the heat release rate fluctuations \cite{poinsot2005theoretical,rayleigh1878explanation}. If the acoustic driving is greater than all the acoustic losses in the system, then the acoustic energy is added to the system \citep{poinsot2005theoretical}. Hence, we use a linear cross-variable correlation for identifying the correlated and uncorrelated dynamics. Further, some of the limitations of linear correlation, when applied to periodic functions are the inability to capture the time lag and the assumption of a linear relation between the variables. Therefore, we use a short window time-delayed correlation as defined in equation \ref{eq: max corr}.}

The regions of ordered activity (ordered regions) show a percolation phase transition in space and time as $\mathrm{Re}$ is increased (Fig.~\ref{Fig: percolation}). The evolution of  {ordered regions}  for the state of chaos, intermittency and limit cycle are shown in figures \ref{Fig: percolation}(a), \ref{Fig: percolation}(b), and \ref{Fig: percolation}(c) respectively. During the state of chaos, ordered regions are small and scattered, and they appear and disappear erratically (Fig.~\ref{Fig: percolation}(a)). However, during the state of intermittency, we observe that some randomly occurring ordered regions grow and form a giant ordered region that spans the entire spatial domain. The giant ordered region formed at an instant ($T-T^\prime$ plane) is shown in Fig.~\ref{Fig: percolation}(b). The giant ordered region formed is not sustained long enough; soon it breaks down and disappears (Fig.~\ref{Fig: percolation}(b)). The formation and the breakdown of the giant ordered region are visible in the slice $X-X^\prime$ of the spatiotemporal domain shown in Fig.~\ref{Fig: percolation}(b). During the state of limit cycle oscillations, we find that ordered activity percolates in space and time (Fig.~\ref{Fig: percolation}(c)). 

In summary, as $\mathrm{Re}$ is increased, we observe a phase transition from a predominantly disordered state where ordered regions appearing are small and scattered, to a dominant ordered state where the regions of ordered activity percolate throughout space and time. 

\section{\label{sec:level6}Critical exponents of the percolation phase transition}
\subsection{Order parameter, susceptibility and inactive interval distribution}
Here, we quantify the characteristics of the phase transition observed in the previous section. The variable $S$ allows us to further quantify the system in terms of the {order parameter} and related statistics. The order parameter ($\rho$) is a measure of the degree of ordered activity in the system dynamics. The order parameter is defined as the fraction of sites depicting order at a given time instant. Formally, it can be expressed as
\begin{equation}
    \rho(t) = \frac{1}{L^2}\sum_{\mathbf{x}} S(\mathbf{x},t),
\end{equation}    
where, $L^2$ is the overall domain size and the summation spans across the entire spatial domain. Consequently, the mean order parameter is expressed as: ${\bar\rho}  = 1/N\sum_t \rho(t)$, where $N$ is the total number of points in the time series. Further, the ensemble average of the order parameter $\langle \bar{\rho} \rangle$ is obtained from ten realizations of the experiment. Here, and in the following, $\langle\cdot\rangle$ implies an ensemble average; refer to Appendix \ref{Appendix: Rec} for more details on the ensemble averaging.

\begin{figure*}[ht]
    \centering
    \includegraphics[width=\linewidth]{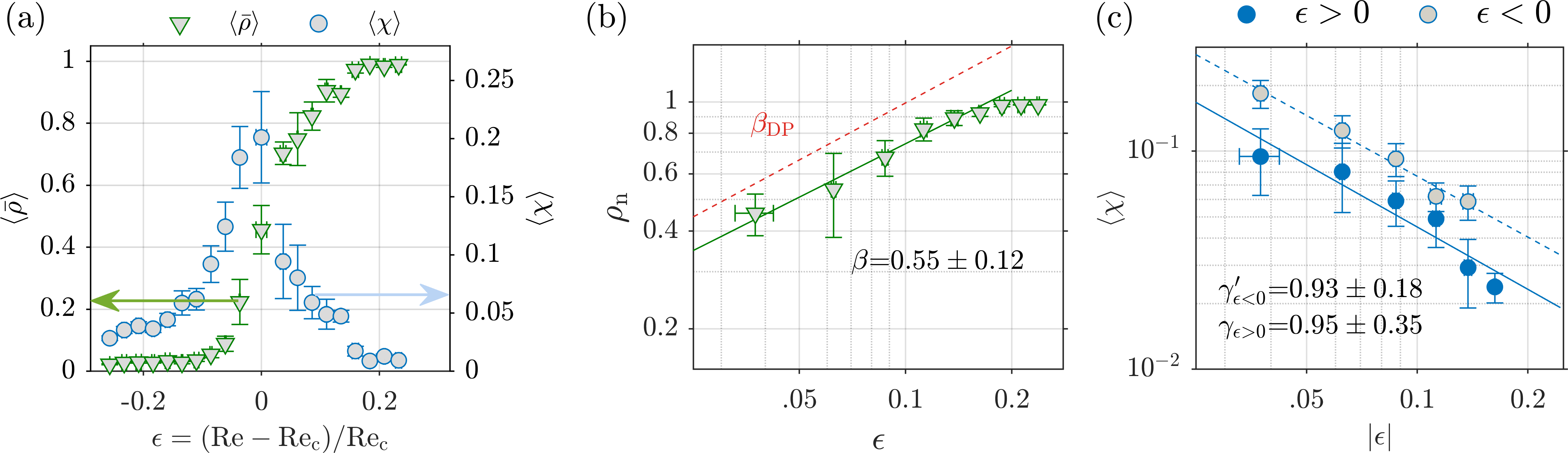}
    \caption{(a) Variation of the order parameter $\langle \bar{\rho} \rangle$ and the susceptibility $\langle \chi \rangle$ as a function of distance from the critical point, $\epsilon={(\mathrm{Re}-\mathrm{Re_c})}/\mathrm{Re_c}$. (b) Scaling of the normalized order parameter $\rho_{\mathrm{n}}$ with an exponent $\beta=0.55 \pm 0.12$. The scaling associated with the universality class of $2+1$ directed percolation ($\beta_{\mathrm{DP}}=0.583$) is also shown by the dashed line as a reference. (c)  {Power law scaling of $\langle \chi \rangle$ as a function of $|\epsilon|$ as the critical point is approached from either side, with the exponents $\gamma^\prime_{\epsilon<0} = 0.93 \pm 0.18$ and $\gamma_{\epsilon>0} = 0.95 \pm 0.35$.} The uncertainty in the scaling exponents is obtained by considering a 90\% confidence; all the error bars correspond to the standard deviation.}
    \label{Fig: OP_X scaling}
\end{figure*}

\begin{figure*}[ht]
    \centering
    \includegraphics[width=\linewidth]{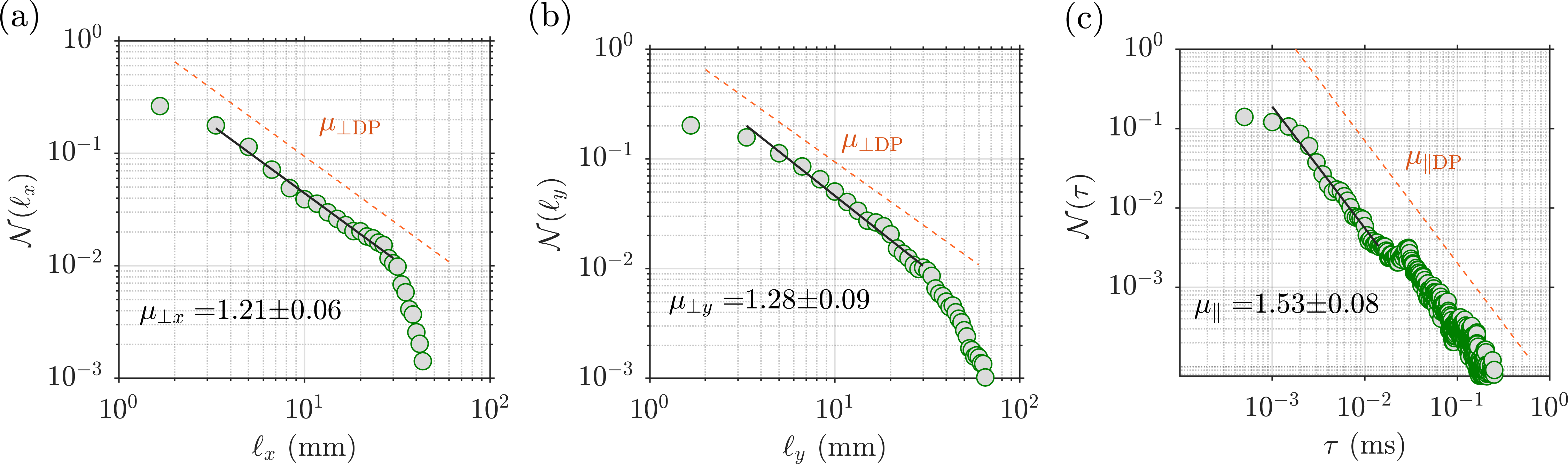}
    \caption{Power law scaling of the probability distribution $\mathcal{N}$ of inactive intervals in (a, b) space ($\ell_x,\ell_y$) and in (c) time ($\tau$) close to the critical point $\epsilon \approx 0$. The observed scaling behavior closely follows the power law scaling associated with the inactive interval distribution of $2+1$ DP (dashed line).} 
    \label{Fig: cluster dist}
\end{figure*}
In real systems, the value of the order parameter at the critical point $\rho_{(\epsilon=0)} \neq 0$ due to the finite size of the system \cite{kiss2002emerging}. Thus, it is important to introduce the normalization $\rho_{\mathrm{n}} = (\langle \bar \rho \rangle - \langle \bar\rho \rangle_0)/(1-\langle \bar\rho \rangle_0)$ such that $\rho_{\mathrm{n}}$ varies between 0 and 1, so as to accurately obtain a power law mentioned as in Eq.~\ref{Eq-Order Parameter Scaling}. Here, $\langle \bar\rho \rangle_0 $ represents the ensemble average of the order parameter at the critical point.  We further define the variance of the order parameter which is referred to as the {susceptibility}, as
\begin{eqnarray}
    \chi=\frac{1}{N} \sum_t \left[\rho^2(t) - (\bar{\rho})^2\right].
\end{eqnarray}
Finally, close to the critical point, the normalized order parameter and susceptibility are well-known to scale as power laws \cite{stanley1971phase}:
\begin{align}
    \rho_{\mathrm{n}} \sim \epsilon^\beta, \qquad \langle\chi\rangle\sim \epsilon^{-\gamma},
    \label{Eq-Order Parameter Scaling}
\end{align}
where, $\beta$ and $\gamma$ are the respective scaling exponents and $\epsilon = (\mathrm{Re}-\mathrm{Re_c})/\mathrm{Re_c}$. \textcolor{black}{Here, $\mathrm{Re_c}$ is identified as the value of the Reynolds number corresponding to the maximum value of susceptibility. The mean value of the critical Reynolds number obtained from an ensemble of ten experiments is $(5.27 \pm 0.04) \times 10^4$. Here, the uncertainty in $\mathrm{Re_c}$ is estimated from the standard deviation value.}


{Figure~\ref{Fig: OP_X scaling} presents the scaling behavior of the order parameter and susceptibility as a function of $\epsilon$}. We find that the ensemble average of the mean order parameter $\langle \bar{\rho} \rangle$ gradually increases from close to $\langle \bar{\rho} \rangle=0$ to $\langle \bar{\rho} \rangle=1$ during the transition (Fig.~\ref{Fig: OP_X scaling}(a)). Between order and chaos, we observe that the variance of the order parameter fluctuations becomes extremely significant and diverges (Fig.~\ref{Fig: OP_X scaling}(a)).  {Such a divergence clearly reveals that the system is approaching a critical point and an impending phase transition.}


\textcolor{black}{Next, we observe that the normalized order parameter $\rho_{\mathrm{n}}$ increases following the power law: $\rho_{\mathrm{n}} \sim  \epsilon^{\beta}$ with the exponent $\beta=0.55 \pm 0.12$ for $\epsilon > 0$ (Fig.~4(b)). However, no such power law scaling behavior is observed for the variation of $\rho_{\mathrm{n}}$ for $\epsilon<0$.} The scaling \textcolor{black}{observed for $\epsilon > 0$} closely follows the scaling behavior of directed percolation. Further, we find that the susceptibility diverges with a power law: $\langle \chi \rangle \sim \epsilon^{-\gamma}$ with the exponents $\gamma=0.95 \pm 0.35$ for $\epsilon > 0$ and $\gamma^\prime= 0.93 \pm 0.18$ for $\epsilon < 0$ respectively (Fig.~\ref{Fig: OP_X scaling}(c)). Thus, the observed transition is continuous with susceptibility exhibiting a diverging behavior between chaos and order.

{Here, and in what follows next, the uncertainty in the value of scaling exponents is obtained by considering the $90\%$ confidence. Further, the vertical and horizontal error bars are calculated based on the standard deviation from the ensemble of the mean order parameter values and $\epsilon$ values within bins of size $\delta  \epsilon=0.025$.}

{Next, we quantify the critical exponents associated with the inactive interval distribution. Inactive interval refers to the duration ($\tau$) or separation ($\ell_x, \ell_y$) between two consecutive ordered activities, as illustrated in Fig.~\ref{Fig: percolation}(b)}. We obtain the probability distribution of inactive interval distribution in space [$\mathcal{N}(\ell_x),~\mathcal{N}(\ell_y)$] and time [$\mathcal{N}(\tau)$] from the percolation diagram, as highlighted in Fig.~\ref{Fig: percolation}. The resulting distributions are plotted in Fig.~\ref{Fig: cluster dist}. We find that close to the critical point of the transition, the distributions of inactive intervals also exhibit power law scaling of the form:
\begin{align}
\mathcal{N}(\ell_x) \sim  \ell_x^{-\mu_{\perp x}}, \enspace \mathcal{N}(\ell_y) \sim  \ell_y^{-\mu_{\perp y}}, \enspace \mathcal{N}(\tau) \sim \tau^{-\mu_{\parallel}}.    
\end{align}


\textcolor{black}{We find that, the scaling exponents for the distribution of inactive intervals along $x$ and $y$ are $\mu_{\perp x}=1.21 \pm 0.06$ and  $\mu_{\perp y}=1.28 \pm 0.09$ in the range of 1.5 mm to 31 mm. The observed scaling behavior is consistent for all the trials of experiments with the exponents nearly remaining the same (refer to Appendix \ref{Appendix: inactive interval}). 
} \textcolor{black}{The distribution of inactive intervals in time exhibits a scaling behavior between 0.001 s to 0.134 s with a scaling exponent of $\mu_{\parallel}={1.53 \pm 0.08}$. We observe a scaling law for three decades for the distribution of inactive time intervals. However, the range of spatial scales where the scaling law is observed is limited by the finite size of the system (the combustor used in our experiments). As a result, we observe only a limited range of scales over which the scaling law is observed in the spatial domain.} 

Power law behaviors in the distribution of inactive intervals signify the scale-invariant nature of the system at the critical point. We also find that the critical exponents corresponding to the inactive interval distributions [$\mathcal{N}(\tau)$, $\mathcal{N}({\ell_x})$ and $\mathcal{N}({\ell_y})$]  fall into the universality class of 2+1 DP (see Appendix \ref{Appendix: inactive interval}).  

 \begin{figure*}[ht]
    \centering    
    \includegraphics[width=\linewidth]{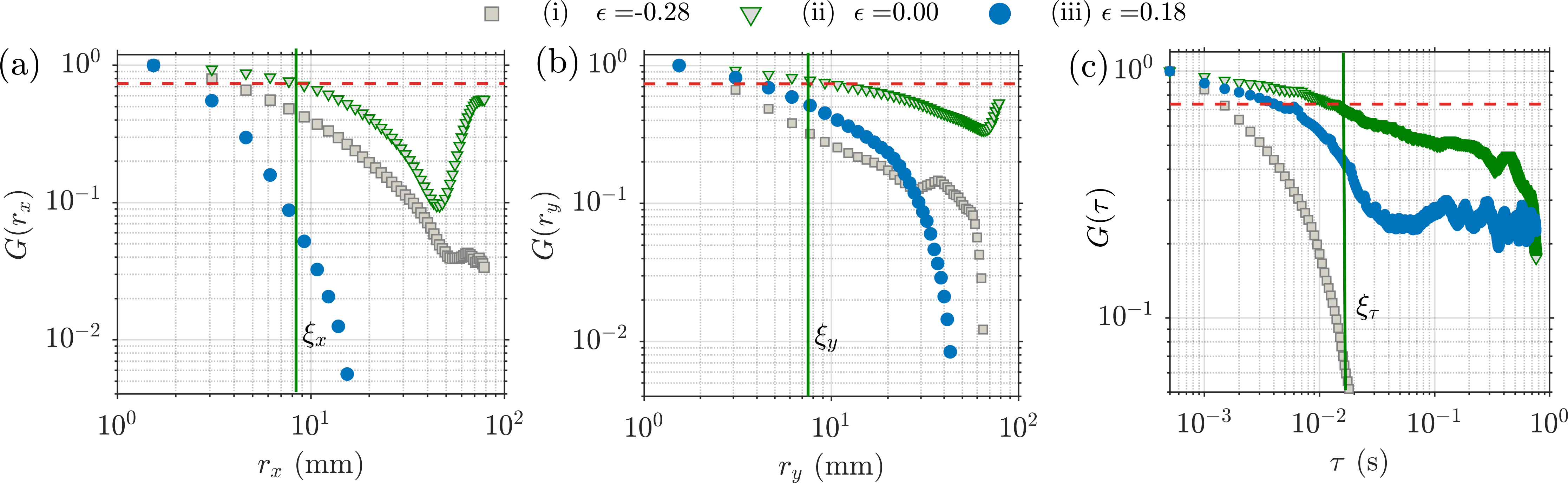}
    \caption{(a-c) Normalized correlation function $(G)$ along $x$, $y$ and $t$ are shown for three $\epsilon$ values, (i) $\epsilon=-0.28$, (ii) $\epsilon=0 $, and (iii) $\epsilon=0.18 $ in the log-log scale. The correlation length ($\xi_{x}$ and $\xi_{y}$ along $x$ and $y$ directions) and the correlation time ($\xi_\tau$) have also been identified as the length or duration where the correlation function crosses $G=2/e$ (dashed line). Correlation $G$ persists the longest at the critical point $\epsilon=0$.}
    \label{Fig: correlation function}
\end{figure*}

\begin{figure*}[t]
    \centering
    \includegraphics[width=\textwidth]{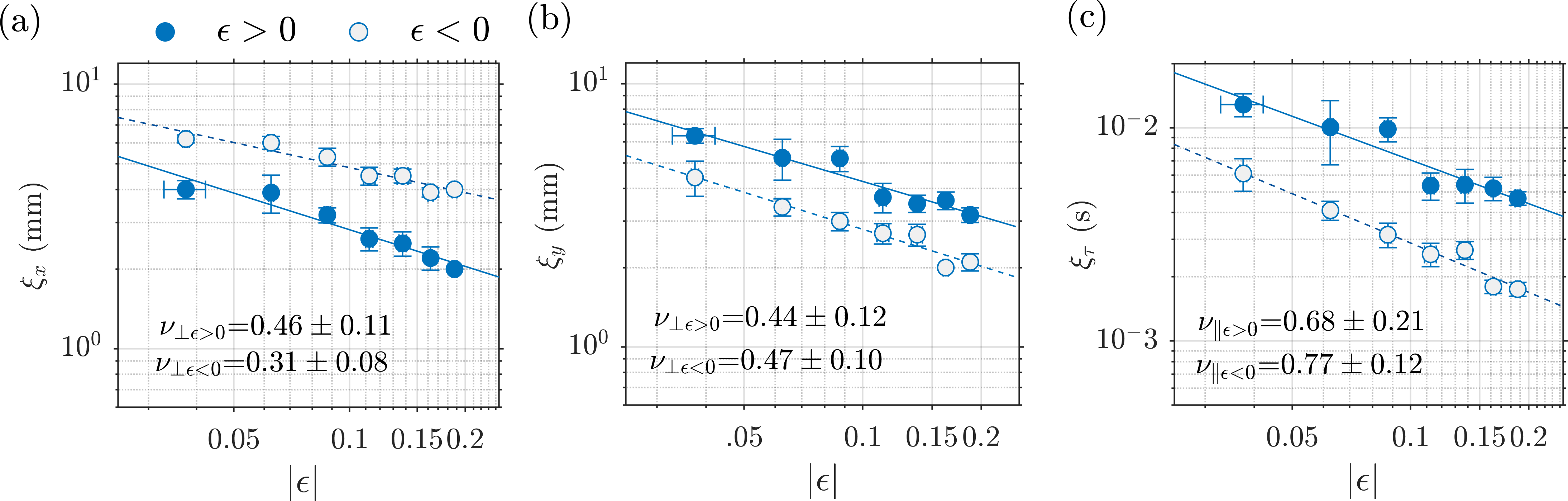}    
    \caption{Log-log plot showing the variation of the (a, b) correlation length ($\xi_{\perp x}$, $\xi_{\perp y}$) and the (c) correlation time ($\xi_{\tau}$) with respect to $\epsilon$. A power law of the form $\xi \sim |\epsilon|^{-\nu}$ is fitted with observed values of the correlation length and time for $|\delta \epsilon| < 0.16$. We observe that the correlation length and duration grow as a power law as the system approaches the critical point from either side of the critical point.}
    \label{Fig: correlation time}
\end{figure*}
\subsection{Scaling of correlation length and time}
In complex systems, significant correlations arise in space and time due to the interaction between the constituent subsystems. We define the two-point correlation function for the scalar field $S(\textbf{x},t)$, which quantifies how the state of the system at two points separated by duration $\tau$ and space $\textbf{r}=[r_x,r_y]^T$ are inter-related.  {Formally, 
 \begin{align}
 g(\tau) &= \langle [S^\prime(\textbf{x},t)S^\prime(\textbf{x},t+\tau)] \rangle,\\
 g(\textbf{r}) &= \langle [S^\prime(\textbf{x},t)S^\prime(\textbf{x+r},t)] \rangle.
 \end{align}
 Here, $S^\prime=S-\langle S\rangle$ and the averaging $\langle f(\textbf{x},t) \rangle = 1/NL^2 \sum_{\mathbf{x},t}f(\textbf{x},t)$ is defined over space and time. The correlation is further normalized by its value at the origin, or the auto-correlation, $g(\tau=0)$, $g(\textbf{r}=0)$ such that we have: $G(\tau) = g(\tau)/g(\tau=0)$ and $G(\textbf{r}) = g(\textbf{r})/g(\textbf{r}=0)$.}

The plot for normalized correlation functions ($G$) along $x,~y$, and $t$ for three different values of $\epsilon$, before the critical point, very close to the critical point, and after the critical point ($\epsilon = -0.28$,  $0$, and $0.18$) are shown in Fig.~\ref{Fig: correlation function}(a-c). As the distance (duration) between the spatial locations (instances) increases, the correlation function decreases.
\textcolor{black}{In the vicinity of the critical point, $G(r_x)$, $G(r_y)$, and $G(\tau)$ decay relatively slower as compared with the cases far from the critical point (see Fig.~\ref{Fig: correlation function}, or refer to Appendix \ref{Appendix: correlation fun}).} {Such persistence and slow decay close to the critical point, imply that the correlation length and time become large due to the scale-invariant nature of clusters of ordered activity arising in the system.}

{The correlation time ($\xi_\tau$) and length ($\xi_{x}$ and $\xi_{y}$ ) are then defined as the time or separation at which the correlation function crosses $G=2/e$ \cite{cavagna2010scale, kesselring2012nanoscale,menard2016comparison} (see Appendix \ref{Appendix: correlation fun}). This is also indicated in Fig.~\ref{Fig: correlation function}(a-c).} The correlation time ($\xi_{\tau}$) or the correlation length ($\xi_{x}$, $\xi_{y}$ along $x$, $y$ respectively) is a measure of how far the correlation persists in the system. The variation in correlation length and correlation time are shown in Fig.~\ref{Fig: correlation time}(a-c). We find that $\xi_\tau$,  $\xi_{x}$ and $\xi_{y}$ increase in a power law manner as we approach the critical point from either side. Thus, for $\epsilon>0$, we obtain the power law exponents for the variation of $\xi_\tau$, $\xi_{x}$ and $\xi_{y}$ as $\nu_\parallel=0.68 \pm 0.21$, $\nu_{\perp x}=0.46 \pm 0.11$, and $\nu_{\perp y}=0.44 \pm 0.12$. Similarly, for $\epsilon <0$, we obtain $\nu_\parallel=0.77 \pm 0.12$, $\nu_{\perp x}=0.31 \pm 0.08$, and $\nu_{\perp y}=0.47 \pm 0.10$ respectively. Such growing correlation time and length are characteristics of a second-order phase transition \cite{stanley1971phase}. \textcolor{black}{The observed differences in the values of $\nu$ along $x$ and $y$ directions is due to the spatial inhomogeneity as observed in the inactive interval distribution (Fig.~\ref{Fig: cluster dist}). The inhomogeneity arises due to the presence of the shear layer, wake regions, and the streamwise ($x$) direction of the mean flow.}

\section{\label{sec:level7}DISCUSSION}

\begin{table}[ht]
\caption{\label{table1}
Critical exponents obtained for the percolation phase transition in the turbulent reactive flow system.}
\begin{ruledtabular}
\begin{tabular}{llll}

\textrm{Exponent}&
\textrm{ 
\begin{tabular}{l}
 Turbulent reactive      \\
 flow system     
\end{tabular}
} &
\textrm{2+1 DP}\footnote{\label{note1}Reference \cite{henkel2008non}}
\\
\colrule
$\beta$ \footnote{\label{note1}For $\beta$, $\nu$, and $\gamma$, the exponents mentioned are obtained for $\epsilon > 0$ }  & $0.55 \pm 0.12$         &  $0.583 $ \\
$\mu_{\parallel}$\footnote{\label{note3}For $\mu$, the exponents are measured for $\epsilon \approx 0$}  & $1.53 \pm 0.08$             & $1.549$  \\
${\mu_{\perp}}^{\text{c,}} $\footnote{\label{note2}For $\mu_{\perp}$ and $\nu_{\perp}$, exponents measured in $x$ and $y$ directions are shown in this order.} & 
$1.21 \pm 0.06$ ($1.28 \pm 0.09$) & {$1.204$} \\
${\nu_{\parallel}}^{\text{b}}$   &   $0.68 \pm 0.21$  &   $1.295$       \\
${\nu_{\perp}}^{\text{b,d}}$ & $0.46 \pm 0.11$   ($0.44 \pm 0.12$)     &  $0.733 $   \\               
$\gamma^{\text{b}}$           & $0.95 \pm 0.35$       &     $0.2998$ 
\end{tabular}
\end{ruledtabular}
\end{table}

We conducted experiments in a confined turbulent reactive flow system and studied the evolution of the correlated dynamics between the flame and the acoustic subsystems. We observe that the correlated dynamics undergoes a percolation phase transition during the emergence of order from chaos. We find that the order parameter gradually increases from zero to one during the percolation phase transition. The correlation time, correlation length, and susceptibility diverge at the critical point of the transition. The critical exponents that characterize this phase transition are listed in Table \ref{table1}. We find that close to the critical point, the critical exponents corresponding to the normalized order parameter ($\beta$), and the distribution of inactive intervals along space ($\mu_{\perp x}$ and $\mu_{\perp y}$) and time ($\mu_{\parallel}$) fall into the universality class of 2+1 DP. We observe that the choice of a threshold value within a range of thresholds does not affect our results (refer to Appendix \ref{Appendix: effect of threshold} for details).  \textcolor{black}{Further, similar scaling exponents can be obtained if the critical Reynolds number is chosen as the Reynolds number where the correlation length is maximum  (refer to Appendix \ref{Appendix: Rec} for details).} Our results suggest that the critical phenomenon we studied belongs to the universality class of DP.

The DP conjecture states that, the transition in a system with a unique absorbing state, short-range dynamic rules, and the absence of special attributes such as the existence of inhomogeneities belongs to the universality class of directed percolation \cite{janssen1981nonequilibrium}. In expt. systems, a pure absorbing state is challenging to achieve due to the inherent fluctuations in the system \cite{hinrichsen2000possible}. Moreover, this limitation makes it hard to identify systems that belong to the universality class of DP through expt. realizations \cite{hinrichsen2000possible}. Despite these difficulties, a handful of expt. studies showed that the transition in turbulent liquid crystals \cite{takeuchi2007directed}, the transition in ferrofluids \cite{rupp2003critical}, the onset of the Leidenfrost effect \cite{chantelot2021leidenfrost}, and the transition from laminar to turbulent state in fluid mechanical systems \cite{lemoult2016directed,sano2016universal} fall under the universality class of DP. 

In our system, the reactant mixture entering the combustion chamber burns while being convected downstream; and the propagation of these reactions (burning reactants) is analogous to a contact process \cite{unni2018flame}. 
The dynamics of the ordered activity are governed by this contact process as well as the fluctuations induced by the global acoustic field. A unique absorbing state is absent during the chaotic state in the turbulent reactive flow system since small regions of ordered activity erratically appear and disappear for $\epsilon < 0$. The inherent fluctuations in the system due to turbulence could be the reason for not observing a unique absorbing state \cite{hinrichsen2000possible}. However, we observe that three critical exponents ($\beta$, $\mu_\perp$, and $\mu_\parallel$) fall into the universality class of DP. The robustness of the universality class of DP  under the relaxation of the DP conjecture \cite{munoz1998phase,munoz1996critical,mendes1994generalized,bhoyar2022robustness} could be why we still observe the critical exponents to be the same as the universality class of DP.

Numerical studies on many spatially extended systems suggest that synchronization transition could belong to the universality class of DP \cite{munoz2003stochastic,jabeen2005dynamic,ginelli2003relationship,ahlers2002critical,grassberger1999synchronization,baroni2001transition,droz2003dynamical}. 
Future studies on identifying the synchronization activity between the flame and acoustic subsystems and its spatiotemporal evolution will help to characterize the synchronization transition in turbulent reactive flow systems.

The universality class of DP typifies phase transition observed in a wide variety of physical systems \cite{Hinrichsen2000}. Further, the universality in statistical models provides insights into how the correlated dynamics (between the subsystems) in diverse systems such as turbulent flow \cite{sano2016universal,lemoult2016directed} and biological systems \cite{carvalho2021subsampled, korchinski2021criticality, cowan2016wilson} are connected, irrespective of their microscopic system details. The statistical analysis based on interdependent fluctuations between the interacting subsystems could help us gain more insights into emergent phenomena associated with nonequilibrium phase transition in complex systems.

\begin{acknowledgments}
We acknowledge Mr. Midhun, Ms. Anaswara, Ms. G Sudha, Mr. Anand and Mr. Tilagraj for their help in conducting the experiments. ST acknowledges support from the Prime Minister Research Fellowship, Government of India. RIS wishes to express his gratitude to the Department of Science and Technology and Ministry of Human Resource Development, Government of India for providing financial support for our research work under Grant Nos. JCB/2018/000034/SSC (JC Bose Fellowship) and SB/2021/0845/AE/MHRD/002696 (Institute of Eminence grant) respectively.
\end{acknowledgments}

\appendix

\section{\label{Appendix: pressure variation} Acoustic pressure variation across the region of interest}
\begin{figure}[ht]
\centering
\includegraphics[width=\linewidth]{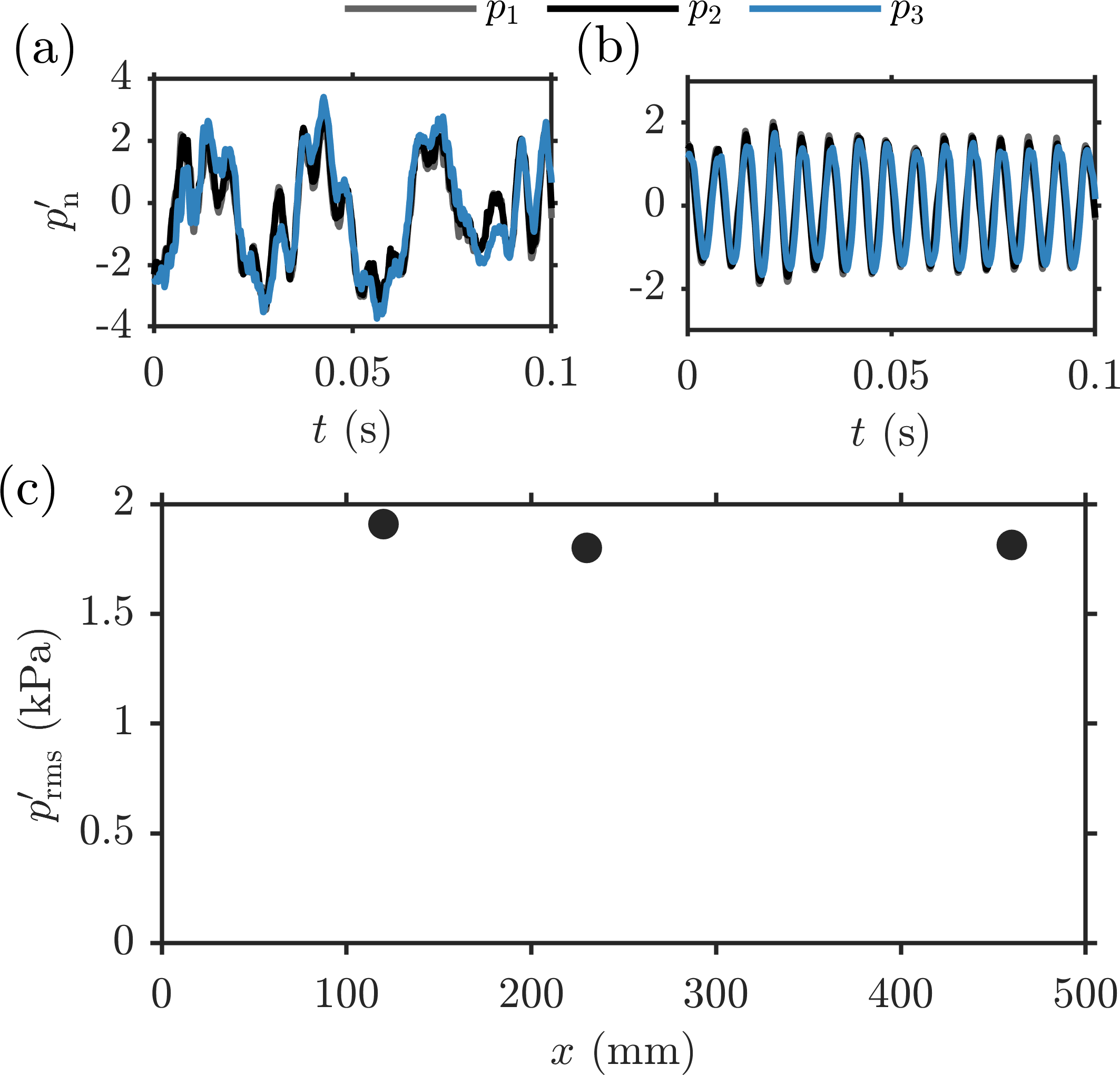}
\caption{ (a, b) The normalized acoustic pressure fluctuations for the state of chaos ($\mathrm{Re}=2.45 \times 10^4$) and limit cycle oscillations ($\mathrm{Re}=3.72 \times 10^4$) at three different locations along the longitudinal axis of the combustor at $x=120$ ($p_1$), 230 ($p_2$), and 460 mm ($p_3$). The normalization is done using their respective standard deviation values.  (c) The root mean square of the acoustic pressure fluctuations for $\mathrm{Re}=3.72 \times 10^4$ for these three different axial locations. We observe that the acoustic pressure fluctuations are uniform within the region of interest.  
}
\label{Fig: pressure spatial}
\end{figure}

We considered that the acoustic pressure variation $p^\prime(t)$ within the region of our study is uniform ($p^\prime(\mathbf{x},t)=p^\prime(t)$). Further, the ordered activity is identified by calculating the cross-correlation ($\mathcal{R}(\mathbf{x},t,\tau_i)$) between the global acoustic pressure ($p^\prime(t)$) and the spatial heat release rate fluctuations ($\dot{q}^\prime(\mathbf{x},t)$). In this section, we show that our consideration of uniform acoustic pressure variation within the region of interest for identifying the ordered activity remains valid. The acoustic pressure fluctuations measured at three locations ($x=120$, $230$, and $460$ mm) along the longitudinal axis of the turbulent reactive flow system are shown in Fig.~\ref{Fig: pressure spatial}. The region of interest for our study is a square section of 80 $\times$ 80 mm. The region of interest is located 50 mm away from the backward-facing step of the combustor along the downstream direction. The pressure fluctuations recorded at these three locations are almost identical. The distance between the locations where the acoustic pressure is measured is greater than the size of the region of interest (80 mm). Hence within the region of interest, the acoustic pressure fluctuations are uniform.

\textcolor{black}{
\section{\label{Appendix: Rec} A different criteria for identifying the critical point}}

\textcolor{black}{The susceptibility and the correlation length diverge at the critical point \cite{yeomans1992statistical,peters2006critical}. In our study, we have selected the critical Reynolds number as the Reynolds number where the susceptibility is maximum. The observed scaling exponents for this criterion are  $\beta =0.55 \pm 0.12$,  $\gamma= 0.95 \pm 0.35$ for $\epsilon > 0$, and $\gamma^\prime=0.93 \pm 0.18$ for $\epsilon <0$.}

\begin{figure}[ht]
    \centering
    \includegraphics[width=\linewidth]{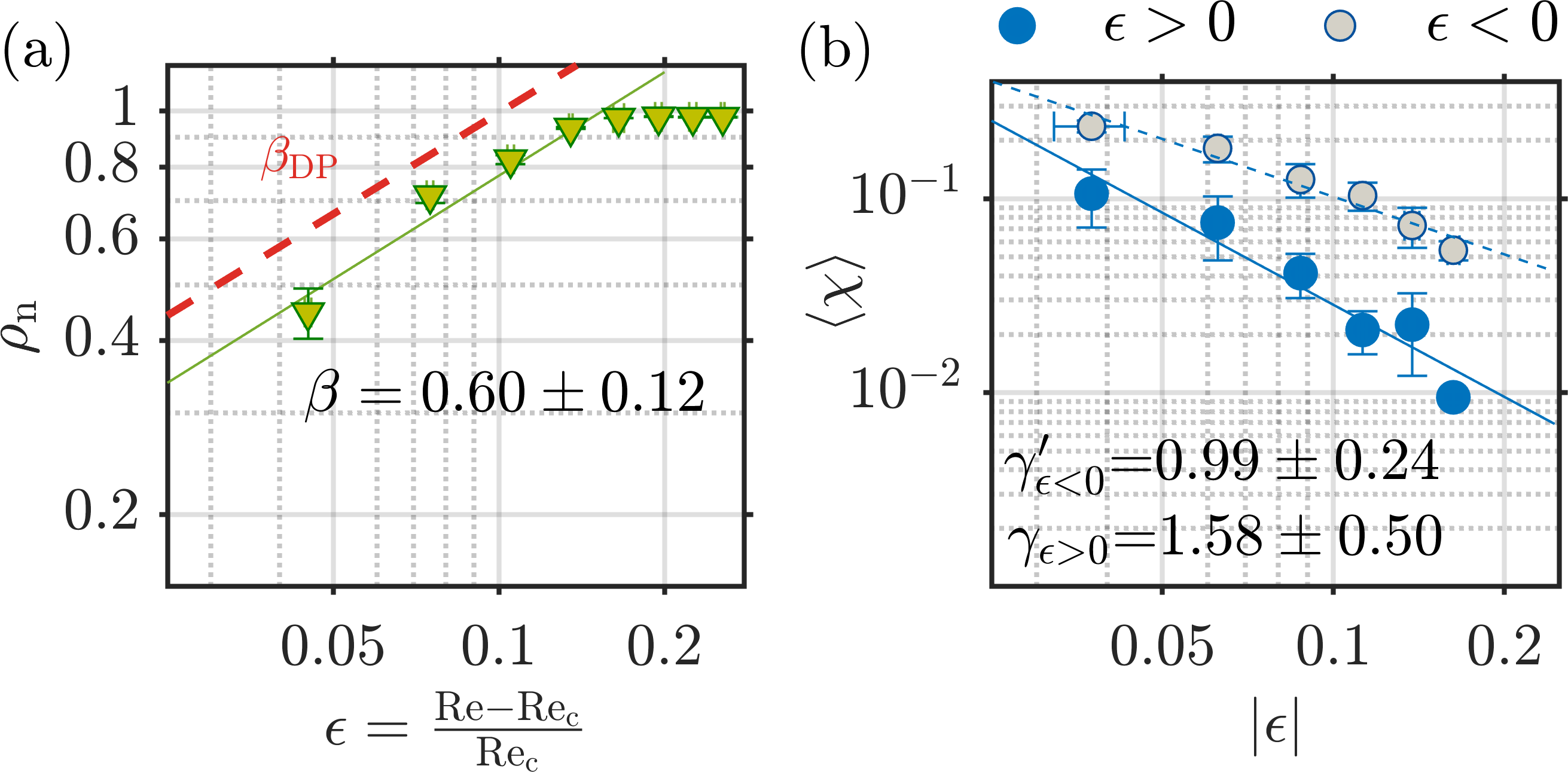}
    \caption{(a, b) The variation of normalized order parameter $\rho_{\mathrm{n}}$ and $\langle \chi \rangle$ with respect to $\epsilon=(\mathrm{Re}-\mathrm{Re_c})/\mathrm{Re_c}$. The critical exponents for the variation of $\rho_{\mathrm{n}}$ is $\beta = 0.6 \pm 0.12$. For the variation of $\langle \chi \rangle$, the obtained power law exponents are $\gamma = 1.58 \pm 0.50  $ for $\epsilon > 0$ and $\gamma^\prime = 0.99 \pm 0.24$ for $\epsilon < 0$.} 
    \label{Fig: critical Re rho X}
\end{figure}

\textcolor{black}{As a different criterion for identifying the critical Reynolds number, we can select the critical point as the control parameter value at which the correlation length is maximum \cite{yeomans1992statistical,peters2006critical}. The variation of the normalized order parameter ($\rho_{\mathrm{n}}$) and $\langle \chi \rangle$ with respect to $\epsilon$ is shown in Fig.~\ref{Fig: critical Re rho X}. The critical exponents obtained using this new criterion are $\beta = 0.6 \pm 0.12$, $\gamma = 1.58 \pm 0.50  $ for $\epsilon > 0$, and $\gamma^\prime = 0.99 \pm 0.24$ for $\epsilon < 0$. The obtained scaling exponents for $\beta$ and $\gamma^\prime$ (for ${\epsilon<0}$) remain nearly the same; however, we observe a variation in the value of $\gamma$ for ${\epsilon>0}$.}

\textcolor{black}{The ensemble average values $\langle\cdot\rangle$ are obtained by considering bins of $\epsilon=( \mathrm{Re}-\mathrm{Re_c})/\mathrm{Re_c}$. Each bin has a width of $\delta  \epsilon=0.025$. }

\begin{figure*}[ht]
    \centering
    \includegraphics[width=\linewidth]{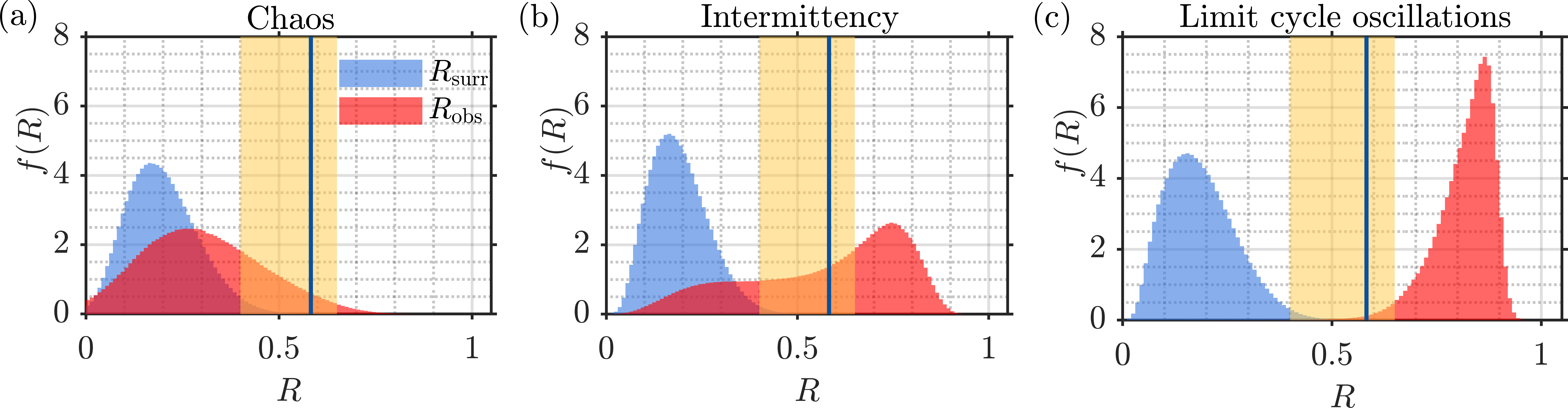}
    \caption{(a-c) The distribution of $R_{\mathrm{surr}}$ and $R_{\mathrm{obs}}$ are shown in blue and red colors for the state of chaos, intermittency, and limit cycle oscillations. The threshold value is selected as the ${99.99}^{\mathrm{th}}$ percentile of the distribution of $R_{\mathrm{surr}}$ from the surrogate test for the state of chaos. The obtained threshold value for identifying the correlated dynamics is 0.58, which is represented by a solid blue line. The percentile for selecting the threshold value is varied from $98^{\mathrm{th}}$ to ${99.999}^{\mathrm{th}}$ percentile, resulting in the variation of threshold value from 0.4 to 0.65, shown as a yellow-colored band. We have shown in Fig.~\ref{FIg: threshold} that the scaling relations nearly remain the same for the threshold values within this range.}
    \label{Fig: Permutation}
\end{figure*}

\section{\label{Appendix: effect of threshold} Surrogate test to choose the correlation threshold}
\textcolor{black}{Here, we use a surrogate test to obtain a suitable threshold for identifying the correlated dynamics. The value of short window time-delayed correlation between the original time series of $\dot{q}_{\mathrm{n}}^\prime ({\mathbf{x}},t)$  and $p_{\mathrm{n}}^\prime(t)$ is denoted as $R_\mathrm{obs}$. For performing surrogate analysis, we find the random permutation of a short window time series of $\dot{q}_{\mathrm{n}}^\prime ({\mathbf{x}},t)$ and find its time-delayed correlation with the corresponding short window $p_{\mathrm{n}}^\prime(t)$ time series (original). The value of correlation thus obtained is denoted as $R_\mathrm{surr}$. In this manner, we perform the random permutation for 1000 times and thus obtain a distribution for the values of $R_\mathrm{surr}$. We consider the original short window time-delayed correlation ($R_\mathrm{obs}$) to be significant if $R_\mathrm{obs}$ is more than the $99.99^{\mathrm{th}}$ percentile of the distribution of $R_\mathrm{surr}$ values. Therefore the $99.99^{\mathrm{th}}$ percentile of the distribution $R_\mathrm{surr}$ value obtained from the surrogate test is selected as the threshold value.}

\begin{figure}[ht]
\centering
\includegraphics[width=\linewidth]{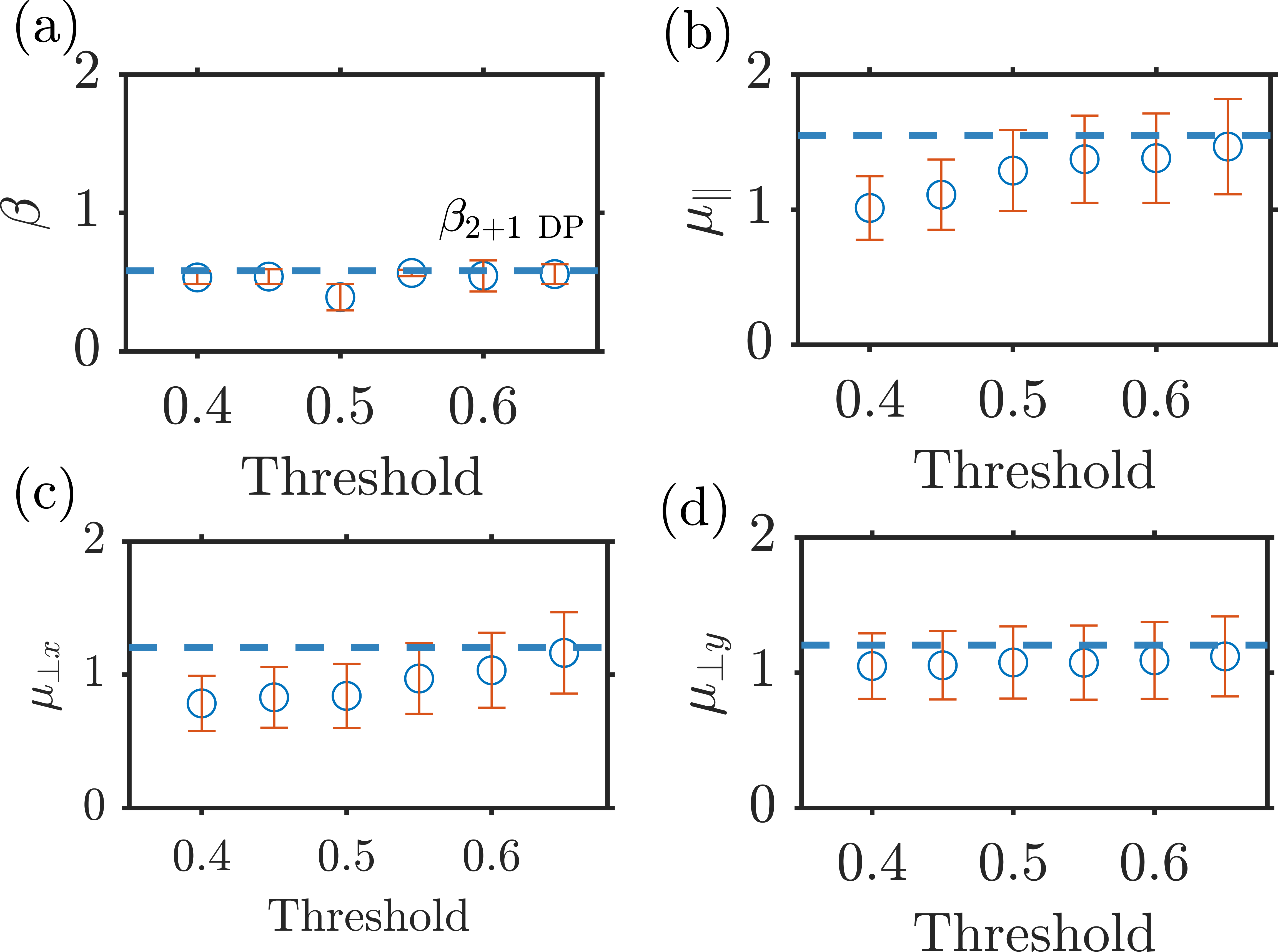}
\caption{ (a-d) The critical exponents $\beta$, $\mu_\parallel$, $\mu_{\perp x}$, and $\mu_{\perp y}$ calculated for various threshold values. The error bars represent the uncertainty in the scaling exponents corresponding to $90\%$ confidence. The horizontal dashed blue lines represent the scaling exponent values of the universality class of 2+1 DP. The reported scaling exponents remain nearly the same for the choice of the threshold values $R_{\text{Th}}$ between 0.4 to 0.65. }
\label{FIg: threshold}
\end{figure}

\textcolor{black}{We have performed the surrogate test for all spatial locations and 2000 different short time windows out of 4000, obtaining the distribution $R_{\mathrm{surr}}$ (blue shaded distribution) for the state of chaos, intermittency, and limit cycle oscillations, as shown in Fig.~\ref{Fig: Permutation}. It is assumed that there is no true correlation between the variables after the random permutation of one of the variables. Therefore the threshold values we obtained for the state of chaos, intermittency, and limit cycle oscillations are 0.582, 0.525, and 0.546 fall in a very narrow range. We choose the threshold from surrogate analysis during chaos so as to differentiate correlated and uncorrelated dynamics during all dynamical states. The threshold value obtained from the surrogate analysis is shown as the solid, blue vertical line in Fig.~\ref{Fig: Permutation}. The distribution of the short window time-delayed correlation between the original time series of $\dot{q}_{\mathrm{n}}^\prime ({\mathbf{x}},t)$  and $p_{\mathrm{n}}^\prime(t)$  ($R_{\mathrm{obs}}$) are shown as red shaded distribution (Fig.~\ref{Fig: Permutation}). During the state of chaos, for the uncorrelated fluctuations between $\dot{q}_{\mathrm{n}}^\prime ({\mathbf{x}},t)$ and $p_{\mathrm{n}}^\prime(t)$, the values of short-window time delayed correlations are relatively low (red shaded distribution in Fig.~\ref{Fig: Permutation}a). On the other hand, for the correlated dynamics between $\dot{q}_{\mathrm{n}}^\prime ({\mathbf{x}},t)$ and $p_{\mathrm{n}}^\prime(t)$ during the state of limit cycle oscillations,  we observe a high value of short-window time delayed correlations with a time delay corresponding to the phase difference between the time series (red shaded distribution in Fig.~\ref{Fig: Permutation}c). The threshold value we have obtained clearly demarcates the correlated and uncorrelated dynamics between  $\dot{q}_{\mathrm{n}}^\prime ({\mathbf{x}},t)$  and $p_{\mathrm{n}}^\prime(t)$  during the state of limit cycle oscillations. Further, we have varied the percentile for the selection of the threshold value from $98^{\mathrm{th}}$ percentile to $99.999^{\mathrm{th}}$ percentile, resulting in the variation of the threshold value from 0.4 to 0.65, which is shown as a yellow-colored band in Fig.~\ref{Fig: Permutation}.
In figure \ref{FIg: threshold}, we have shown that the scaling behavior remains and the values of $\beta$, $\mu_\perp$, and $\mu_\parallel$ are very close to that obtained in table \ref{table1} for threshold values ranging from 0.40 to 0.65. The results presented in our manuscript correspond to the threshold value of $R_{\mathrm{Th}}=0.60$ obtained from the surrogate analysis.}

\begin{figure*}[t]
\centering
\includegraphics[width=0.9\linewidth]{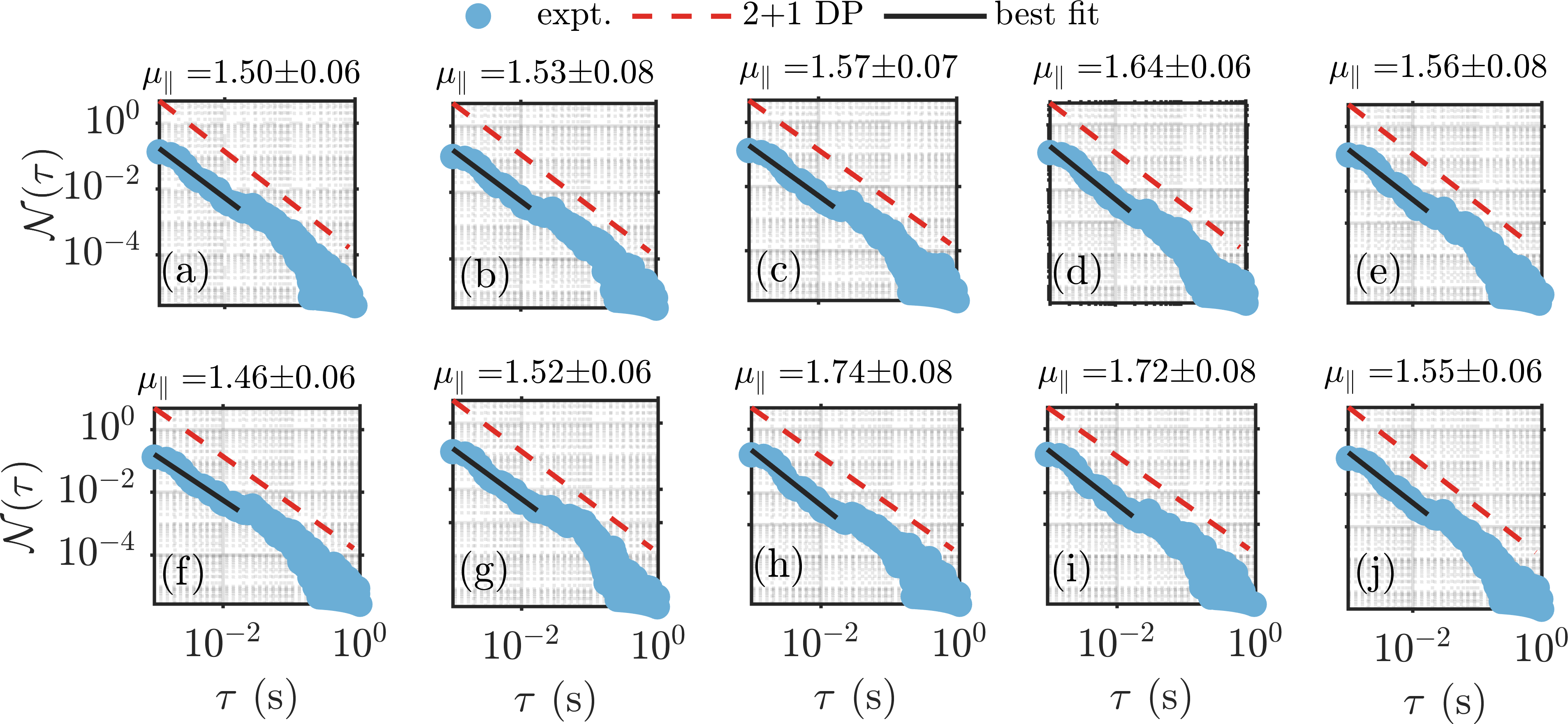}
\caption{The distribution of inactive interval in time ($\mathcal{N}(\tau)$) exhibits power law behavior very close to that of the universality class of 2+1 directed percolation. (a-j) $\mathcal{N}(\tau)$ at the critical point for each trial of the experiments. The dashed line is the reference line corresponding to the universality class of 2+1 directed percolation with a slope of $\mu_{\parallel \mathrm{DP}}=1.5495$. The black line segment is the best fit for the distribution of the $\tau $ values falling within the line segment. There is a close agreement between the best fit for the inactive interval distribution and the universality class of 2+1 DP. The ensemble mean of the exponent corresponding to the best fit ($\mu_\parallel$) from the experiments is $\mu_{\parallel \mathrm{expt}} =1.57$ with a standard deviation of 0.09.}
\label{FIg: clust time}
\end{figure*}

\begin{figure*}[ht]
\centering
\includegraphics[width=0.9\linewidth]{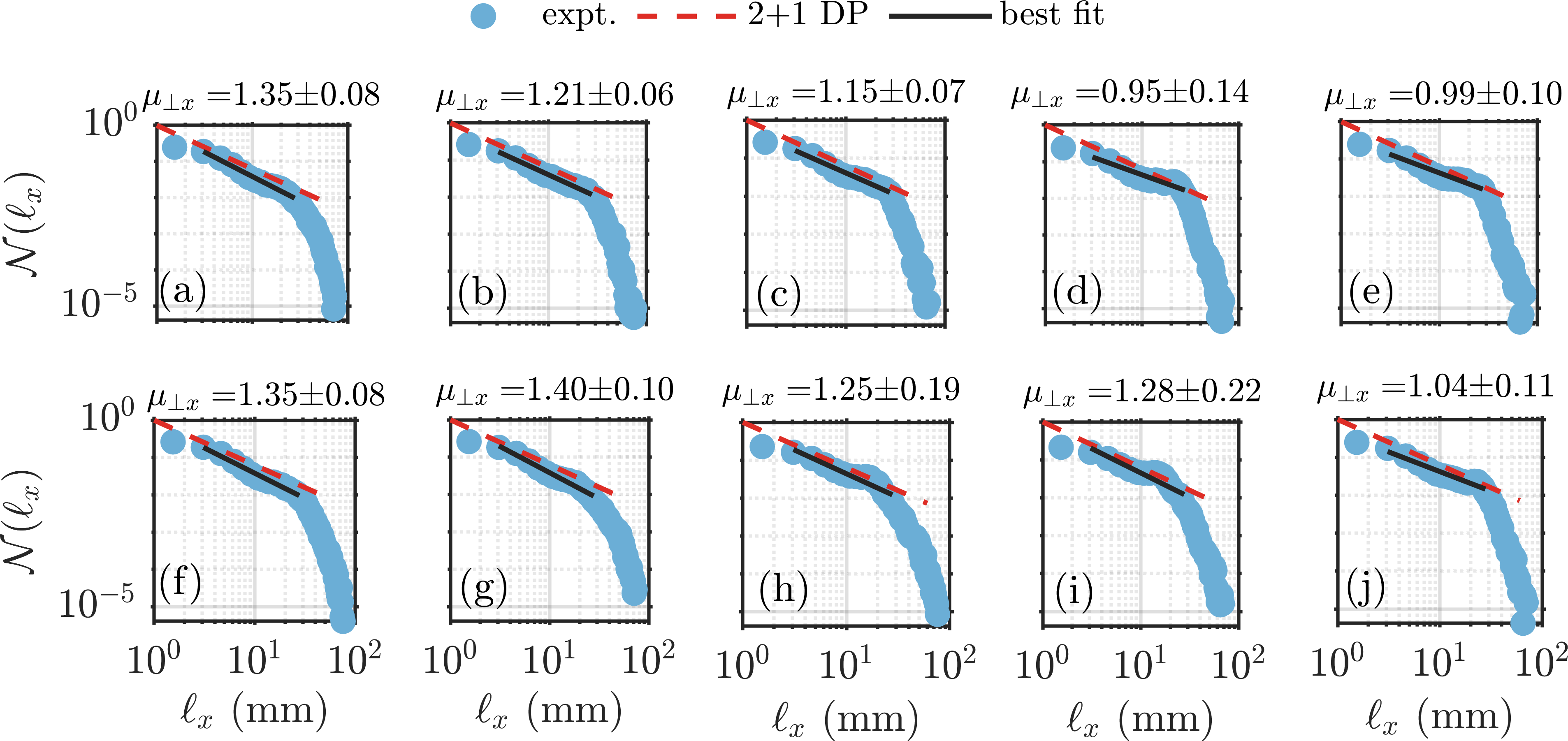}
\caption{The distribution of inactive interval distance ($\mathcal{N}(\ell_x)$) along the spatial dimension ($x$) exhibits power law behavior very close to that of the universality class of 2+1 directed percolation. (a-j) $\mathcal{N}(\ell_{x})$ at the critical point for each of the experiments. The dashed line in (a)-(j) is the reference line corresponding to the universality class of 2+1 DP with a slope of $\mu_{\perp \mathrm{DP}}=1.204$. The black line segment is the best fit for the distribution of the $\ell_x $ values falling within the line segment. The ensemble mean of the scaling exponents for the distribution of inactive interval is $\mu_{\parallel \mathrm{expt}} =1.28$ with a standard deviation of 0.16.}
\label{FIg: clust lx}
\end{figure*}

\begin{figure*}[ht]
\centering
\includegraphics[width=0.9\linewidth]{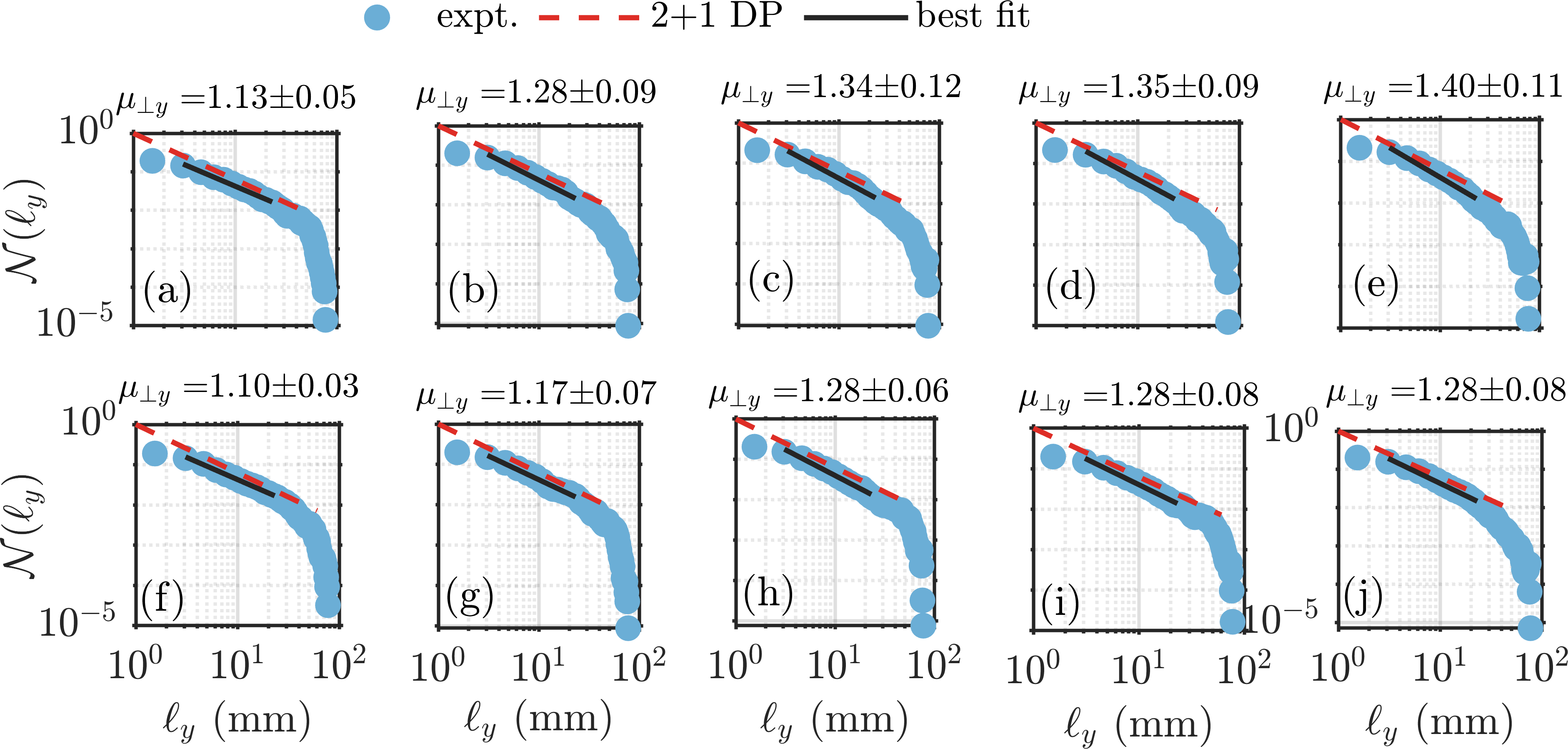}
\caption
{(a-j) The distribution of inactive interval along $y$ direction ($\mathcal{N}(\ell_{y})$) at the critical point for each trial of the experiments. The dotted line in (a)-(j) is the reference line corresponding to the universality class of 2+1 DP  with a slope of $\mu_{\perp \mathrm{DP}}=1.204$. The black line segment is the best fit for the distribution of the $\ell_y $ values falling within the line segment. The ensemble mean of the scaling exponents is $\mu_{\parallel \mathrm{expt}} =1.26$ with a standard deviation of 0.1.}
\label{FIg: clust ly}
\end{figure*}

\begin{figure*}[ht]
\centering
\includegraphics[width=\linewidth]{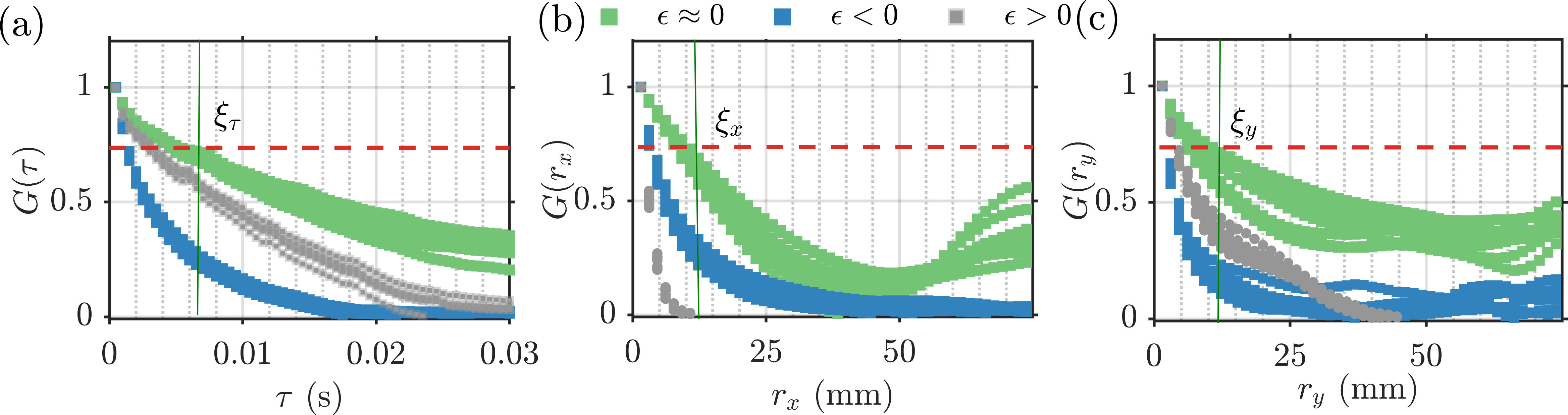}
\caption{(a-c) Normalized correlation function ($G(\tau)$, $G(r_x)$, and $G(r_y)$) along time, $x$, and $y$ directions are shown for ten experiments corresponding to the control parameter values $\epsilon \approx 0$, $\epsilon > 0$, and $\epsilon <0$. The correlation function persists for a longer duration or distance close to the critical point in comparison with the control parameter values away from the critical point.}
\label{Fig: G_xyt_sup}
\end{figure*}
\newpage

\section{\label{Appendix: inactive interval} Scaling behavior of the distribution of inactive intervals}

The distribution of interval between two consecutive ordered activities exhibits a scaling behavior close to the critical point. In this section, we show that the distribution of these intervals is closely matching with that of the universality class of directed percolation. 

At the critical point, the distributions of inactive intervals are characterized by power laws with exponents $\mu_{\parallel}$ in time and $\mu_{\perp}$ in space. The inactive interval distributions along time, $x$ and $y$  for all the trials of the experiment are shown in figures \ref{FIg: clust time}, \ref{FIg: clust lx} and \ref{FIg: clust ly}. The power law behavior observed for the inactive interval distribution is very close to that of the universality class of 2+1 DP across all the trials of the experiment (Fig.~\ref{FIg: clust time}, \ref{FIg: clust lx} and \ref{FIg: clust ly}).

\section{\label{Appendix: correlation fun}Correlation function}
Close to the critical point, the correlations between the fluctuations of the acoustic pressure and the heat release rate persist longer. The correlation function along time and space is defined in equations (8) and (9). The correlation function along time, $x$ and $y$ directions for each experiment corresponding to three conditions $\epsilon \approx 0$, $\epsilon > 0$, and $\epsilon < 0$ are shown in Fig.~\ref{Fig: G_xyt_sup}.  The correlation function close to the critical point exhibits large variations as evident from the green color lines. Further, in all the experiments we have performed, the correlation function decays slowly near the critical point when compared to control parameter values away from the critical point (Fig.~\ref{Fig: G_xyt_sup}). The distance (or duration) at which the correlation function decays to 2/e is defined as the correlation length (or time). The factor 2/e, instead of 1/e is selected owing to the large variations in the correlation function. We observe that, close to the critical point, the correlation function persists for a longer distance (or duration).

\section{\label{Appendix: Phase space reconstruction} Optimum time delay and suitable embedding dimension for phase space reconstruction}
\begin{figure*}[t]
    \centering
    \includegraphics[width=0.9\linewidth]{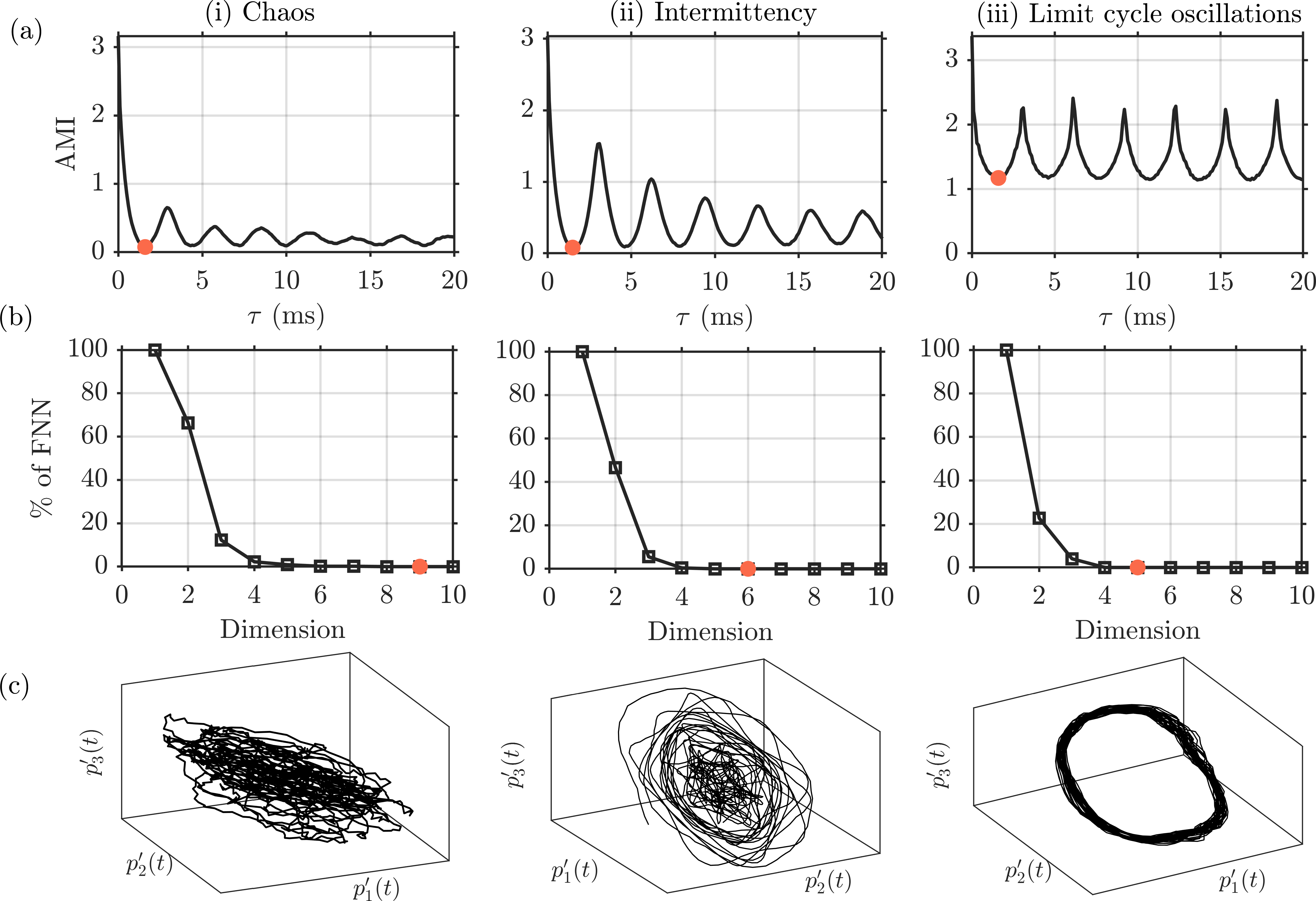}
    \caption{(a) Average mutual information with respect to delay, (b) percentage of FNN with respect to the embedding dimension, and (c) the reconstructed phase space for the state of (i) chaos, (ii) intermittency, and (iii) limit cycle oscillations (LCO). The optimum lag obtained for the state of chaos, intermittency, and LCO are 1.6, 1.5, and 1.6 ms respectively and they are highlighted. The optimal embedding dimensions obtained for the state of chaos, intermittency, and LCO are 9, 6, and 5 and they are highlighted.} 
    \label{Fig: phase space}
\end{figure*}

For reconstructing the phase space, the optimum time delay is selected as the first local minima of the average mutual information \cite{fraser1986independent} and the suitable embedding dimension is calculated using the false nearest neighbor method \cite{kennel1992determining}. A false nearest neighbor (FNN) in the phase space changes its relative position as the embedding dimension is increased. In this method, we keep track of the fraction of the FNN in the phase space as the embedding dimension is progressively increased. When the percentage of FNN falls to zero for a particular value of the embedding dimension ($d-1$) for the first time, the dimension corresponding to the next higher value is considered as the optimal embedding dimension ($d$) required for the phase space reconstruction.  
The optimum lag ($\tau$) and embedding dimension for reconstructing the phase space for the state of chaos, intermittency, and limit cycle oscillations are shown in Fig.~\ref{Fig: phase space}. 
The optimum lag values obtained for the state of chaos, intermittency, and limit cycle oscillations (LCO) are 1.6, 1.5, and 1.6 ms respectively. The optimal embedding dimensions obtained for the state of chaos, intermittency, and LCO are 9, 6, and 5. The percentage of FNN remains the same if the embedding dimension is greater than the optimal dimension ($d$) (Fig.~\ref{Fig: phase space}).

\newpage

\bibliography{apssamp}

\end{document}